\documentclass[pre,aps,superscriptaddress,twocolumn]{revtex4-1}
\usepackage{graphicx} 
\usepackage{subfigure} %
\usepackage{dcolumn}
\usepackage{bm}
\usepackage{tabularx}
\usepackage[latin1]{inputenc}
\usepackage[usenames,dvipsnames]{color}
\usepackage{amsmath}
\usepackage{amsfonts}
\usepackage{mathrsfs}
\usepackage{amssymb} 
\def\be{\begin{equation}}
\def\ee{\end{equation}}

\def\ba{\begin{array}}
\def\ea{\end{array}}
\def\bea{\begin{eqnarray}}
\def\eea{\end{eqnarray}}

\begin{document}

\title{Bending transition in the penetration of a flexible intruder\\ in
  a 2D dense granular medium}

\author{Nicolas Algarra}
\affiliation{Laboratoire PMMH, UMR 7636 CNRS/ESPCI Paris/PSL Research University/Sorbonne Universit\'es, UPMC Univ Paris 06,/Univ. Paris 7 Diderot, 10 rue Vauquelin, 75231 Paris cedex 05, France}

\author{Panagiotis G. Karagiannopoulos}
\affiliation{Laboratoire PMMH, UMR 7636 CNRS/ESPCI Paris/PSL Research University/Sorbonne Universit\'es, UPMC Univ Paris 06,/Univ. Paris 7 Diderot, 10 rue Vauquelin, 75231 Paris cedex 05, France}

\author{Arnaud Lazarus}
\affiliation{Sorbonne Universit\'es, UPMC Univ Paris 06, CNRS, UMR 7190, Institut Jean Le Rond d'Alembert, F-75005, Paris, France}

\author{Damien Vandembroucq}\affiliation{Laboratoire PMMH, UMR 7636 CNRS/ESPCI Paris/PSL Research University/Sorbonne Universit\'es, UPMC Univ Paris 06,/Univ. Paris 7 Diderot, 10 rue Vauquelin, 75231 Paris cedex 05, France}

\author{Evelyne Kolb}
\affiliation{Laboratoire PMMH, UMR 7636 CNRS/ESPCI Paris/PSL Research University/Sorbonne Universit\'es, UPMC Univ Paris 06,/Univ. Paris 7 Diderot, 10 rue Vauquelin, 75231 Paris cedex 05, France}

\date{\today}

\begin{abstract}
We study the quasi-static penetration of a flexible beam in a
two-dimensional dense granular medium lying on an horizontal
plate. Rather than a buckling-like behavior we observe a transition
between a regime of crack-like penetration in which the fiber only
shows small fluctuations around a stable straight geometry and a
bending regime in which the fiber fully bends and advances through
series of loading/unloading steps.
We show that the shape reconfiguration of the fiber is controlled by a
single non dimensional parameter: $L/L_c$, the ratio of the
length of the flexible beam $L$ to $L_c$, a bending elasto-granular
length scale that depends on the rigidity of the fiber and on the
departure from the jamming packing fraction of the granular medium.
We show moreover that the dynamics of the bending transition in the
course of the penetration experiment is gradual and is accompanied by
a symmetry breaking of the granular packing fraction in the vicinity of the
fiber. Together with the progressive bending of the fiber, a cavity
grows downstream of the fiber and the accumulation of grains upstream
of the fiber leads to the development of a jammed cluster of
grains. We discuss our experimental results in the framework of a
simple model of bending-induced compaction and we show that the rate
of the bending transition only depends on the control parameter $L/L_c$.

\end{abstract}

\maketitle

\section{Introduction}

Slender structures are extremely flexible and get easily unstable. The
ratio of the bending stiffness of a beam to its axial stiffness in
tension or compression scales with $(t/L)^2$, the square of the ratio
of the thickness to the length of the beam~\cite{Audoly-book2010}. Similarly, over a
threshold compressive strain of order $(t/L)^2$, the beam undergoes a
buckling instability~\cite{Bazant-book2010}. As a consequence of their high geometrical aspect ratio, slender
structures can be sensitive to low forces of various origins and
exhibit complex mechanical behaviors due to the associated couplings.
This is in particular the case of the elasto-capillary problems
(interactions between slender structures and capillary forces) which
have raised a growing interest in the recent
years~\cite{Roman-JPCM10}.

In the same spirit, the effect of flexibility on fluid-structure
interactions has recently motivated a growing number of studies at low~\cite{Shelley-ARFM11,Lindner-ARFM17} or high Reynolds number ~\cite{Ramananarivo-PNAS11}. Model
experiments of reconfigurations have been performed with flexible fibers (1D), plates (2D) or assembly of plates organized in a circular
pattern (3D) that were placed initially
perpendicular to the incoming flow of air \cite{Gosselin-JFM10}, water \cite{Boudaoud-JFM06} or 2D soap film
 \cite{Alben-Nature02}. In all these cases, the flexible body
experiences shape reconfiguration with self-streamlining and reduction
of the surface area exposed to the flow, thus resulting in a drastic
reduction of the drag force exerted on the object \cite{Vogel-book94}. An immediate field
of application of these problems can be found in bio-physical domains, when a passive or an active elastic
part of a body interacts with the flow~\cite{Coq-PF08,Ramananarivo-PNAS11,XiaoChen-JRSocInterf13,Reis-PRL15}. In particular, in animal locomotion, the flexibility of wings or fins intervene in the flying of birds and insects or the swimming of fishes or eels. Even in micro-organisms the propulsion through flagella takes advantage of the flexibility. Numerous examples can also be found in the plant domain~\cite{DeLangre-CRM12} where the flexibility of cereal stems, tree's trunk, branches or leaves can be beneficial under wind flow for reducing lodging. Even marine algae and plants in aquatic canopies adapt their shape under water current~\cite{Nepf-aquatic11,Barsu-POF16}.

Here we consider an original case of fluid structure interaction
between a slender structure and a granular medium. More specifically,
we study the quasi-static penetration of a flexible beam in a
two-dimensional dense bidisperse mixture of discs lying on a
plate. Applications of this problem range from nuclear
engineering~\cite{Buster-NED16} to biology: growth of
roots ~\cite{Kolb-PlantSoil12,Wendell_ExpMech_2012} in structured soils or locomotion of worms or sandfish lizards in granular media~\cite{HosoiGoldman-ARFM15}.

The complex behavior of granular flows has recently been studied through its interaction with rigid intruders~\cite{Albert-PRL99,Albert-PRE01,Stone-PRE04,Dauchot-PRL09,Dauchot-PRE10,Zippelius-GM12,Kolb-PRE13,Kolb-GM14,Gondret-PRE13,Gondret-PRE16,Behringer-PRE16,LopezdelaCruz-JFM16,Jop-PRE16}. The high sensitivity of flexible intruders to low forces opens here a promising new way of probing the mechanical behavior of  granular media~\cite{Holmes-EML16,Holmes-arxiv17,Kolb-EPJWoC17}. 
The complexity of such a question of \emph{elastogranular mechanics}
(as recently coined in Ref.~\cite{Holmes-arxiv17}) will naturally
arise from the coupling of the low rigidity of the elastic beam with
the discrete nature of the granular flow, its non-local
rheology~\cite{Bouzid-EPJE15} and the emergence of turbulent-like
fluctuations~\cite{Radjai-PRL02,Atman-AIP13,Mizuno-SM16} in the
vicinity of the jamming transition.

In contrast with two recent works
focusing on buckling~\cite{Holmes-EML16,Holmes-arxiv17}, the present
study will highlight a bending transition of the flexible intruder and
an associated transition in the structure of the granular medium.

In the following we first give in section~\ref{setup} a brief
description of the experimental set-up and tools of analysis; we then
discuss in section~\ref{phenomenology} the phenomenology of the
interaction between an elastic beam and a granular flow, we report our
experimental observation of bistability between two different regimes
(straight or bent fiber), we identify the control parameters
(rigidity of the fiber and packing fraction of the granular medium) and propose
a simple model ; in section~\ref{transition} we give a specific focus
on the development of the bending transition and show that it is
associated with a clear symmetry breaking of both the geometry of the
fiber and the density of the granular material in the vicinity of the
fiber; we then propose in section~\ref{model} a model of bending
induced compaction that allows us to reasonably reproduce the
evolution of the fiber deflection in the course of the bending
transition; a summary of our main results is finally given in
section~\ref{conclusion}.

\section{Experimental set-up\label{setup}}

\subsection{The granular medium}

\begin{figure}
\begin{center}
  \includegraphics[width=0.95\linewidth]{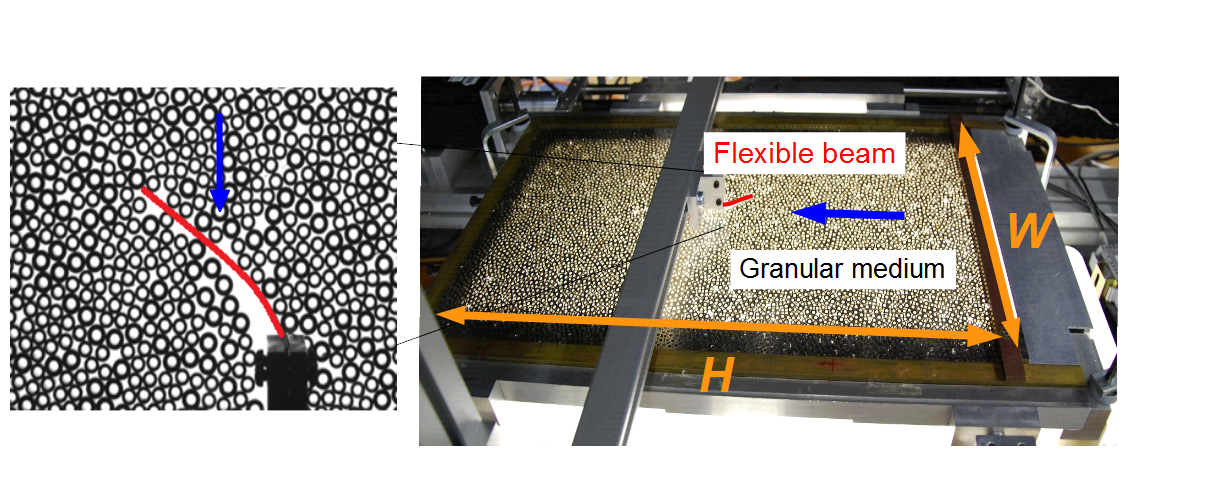}
  \end{center}
\caption{Experimental set-up (right panel): the granular layer is contained in an horizontal rectangular plate of width $W$ and length $H$ that moves at a constant speed along the direction indicated by the blue arrow ($Y$-axis). Bending of the fiber (left panel) produced by the granular flow as observed on a zoom of a picture taken with a CCD camera placed above the setup. The fiber clamping point is fixed in the laboratory frame.}
\label{set-up}
\end{figure}

The 2D granular material is a bidisperse mixture of brass cylinders,
in equal mass proportion, forming a dense and disordered assembly of
rigid disks~\cite{Kolb-PRE13,Kolb-GM14}).  The two kinds of cylinders
of respective outer diameters $d_{1}$~=~4~mm (in number $N_{1}$) and
$d_{2}$~=~5~mm (in number $N_{2}=\frac{4}{7}N_{1}$) lie on a horizontal
glass plate delimited by four brass walls forming a rectangular frame
of width $W$~=~269.5 mm (54~$d_{2}$ along the $X-$axis) and adjustable length
$H$ along the $Y-$axis (Fig.~\ref{set-up}). The total number of cylinders is kept constant
around 6800, but as we varied $H$ in the range 457.5 mm (91 $d_{2}$)
to 470.5 mm (94 $d_{2}$), we could adjust the total
available cell surface and therefore the packing fraction $\phi$. This
packing fraction has been chosen in a small range just below the
assumed jamming packing fraction for our 2D granular medium,
$\phi_{J}$=83.56\%~\cite{Kolb-PRE13}.

\subsection{The flexible beam}

The flexible beam is cut into a mylar sheet of thickness
$t$=350~$\mu$m, Young modulus $E$=3.8~GPa and Poisson ratio $\nu$=0.4  \cite{Boudaoud_Couder_Nature_2000_Mylar}. The flexible
part of the intruder is a long beam, that we will call further fiber,
of length $L$ with a rectangular cross-section defined by the sheet
thickness $t$ and a height $h$~=~3~mm corresponding to the height of the cylinders of the granular material. One of the
intruder extremity is clamped into a rigid metallic part supported by
a transverse arm along the $X$ plane (Fig.~\ref{set-up}).

The anchoring point of the intruder is fixed in the laboratory frame
and is located at an equal distance of the two lateral walls
delimiting the space for the granular layer. The basis of the fiber
is fixed slightly above the bottom glass plate, so that there is no
friction with the bottom. The measurement of the grain-grain friction coefficient gives $\mu_{gg}$=0.32, while the one for the
frictional contact of grains with the glass bottom leads to
$\mu=\mu_{gb}$=0.49$\pm$0.09. At the beginning of the experiment, the
fiber is straight along the $Y-$axis with no contact with neighboring
grains. The anchoring point of the fiber is initially located at a
distance of 80~mm=16~$d_{2}$ (or 120 mm=24~$d_{2}$ depending
on the batch of experiments) from the back wall, to avoid boundary
effects.

\subsection{Principle of the experiment}

A typical experiment consists in translating the granular material
supported by the glass plate along the $Y-$axis against the free
extremity of the fiber. The plate velocity $V_0$ is held constant. In
the plate frame this is equivalent of having a fiber penetrating the
granular material at a constant velocity $V_0$. The experiment is
stopped after a plate displacement of typically 260~mm (52~$d_{2}$).

In this work, we used different packing fractions ranging from
$\phi$=79.0\% to 83.1\% and different fiber lengths between $L$=1~cm and $L$=5~cm.
The velocity of the plate was always kept constant at $V_0$ =5/6 mm/s. This value corresponds to a quasi-static regime in which the forces are dominated by frictional contact forces. Actually we observe that the drag force experienced by
a rigid cylindrical intruder does not depend on velocity in these
ranges of values. A CCD camera of 1600*1200 pixels placed above the
set-up records images at a frequency of 2~Hz, corresponding to a plate
displacement of $U_0$ =5/12~mm=$d_2/12$ between 2 successive
images. Ten similar experiments are performed for each set of
parameters $L$ and $\phi$, and grains are carefully remixed between
two consecutive experiments and fiber checked and replaced if
necessary.

\subsection{Segmentation and image processing}

For all images of each batch of experiments, a technique of
correlation on gray levels has been used for determining the grain
centers with a sub-pixel accuracy. Then the grains displacements from
one image to the next one have been computed. Furthermore, the full
deflected shape of the fiber is obtained for most images. The
delicate challenge for image analysis was to locate the very thin and
elongated fiber amongst the circular grains surrounding it. For this
purpose, the gray-level image has been convoluted with different
filters to enhance the contrast of the fiber and disconnect it from
the neighboring grains. This first step gave a skeleton of the fiber,
which was further fully reconstructed by starting from the fiber
anchoring point. From the fiber shape, different informations can be
obtained like the free end lateral deflection $\delta$ or the local
slope $\theta$ of the fiber relative to the $Y-$axis. This local slope
can be computed as a function of the curvilinear abscissa $s$
(normalized by the length of the fiber), with $s$=0 at the clamped
extremity and $s$=1 at the free extremity.


\begin{figure}
\begin{center}
(a)\includegraphics[width=0.85\linewidth]{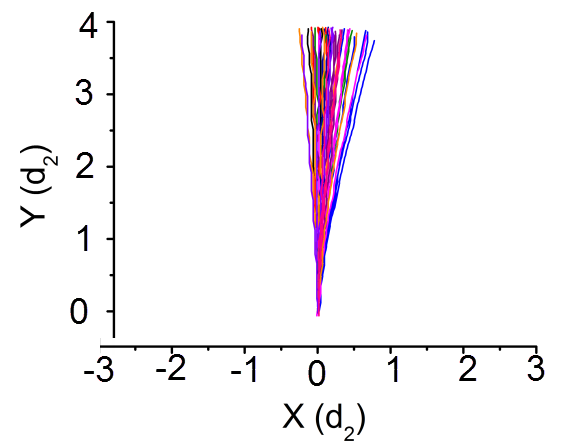}	
(b)\includegraphics[width=0.85\linewidth]{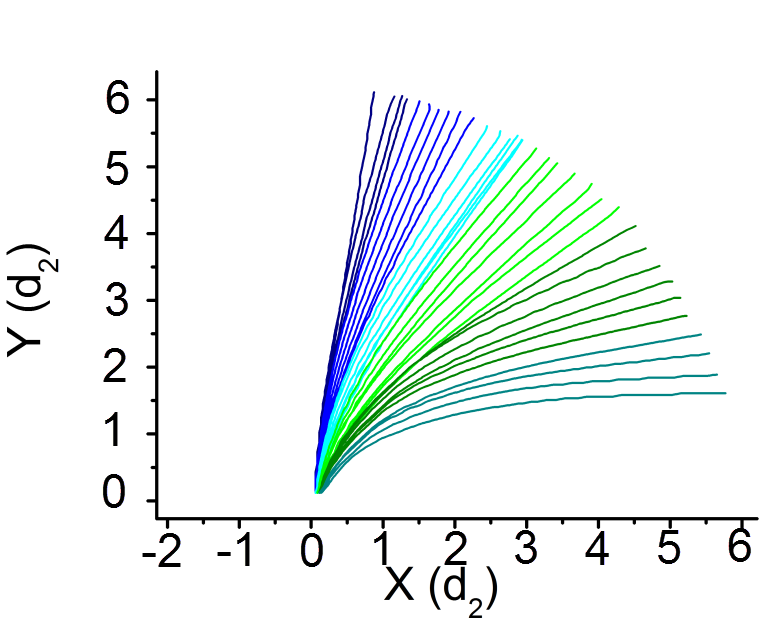}
\end{center}
\caption{(a) Fluctuations of an elastic beam of length $L=2$ cm in a granular medium of packing fraction $\phi=80.94\%$ (jiggling regime). The clamping point of the fiber is located in (0,0). The granular flow direction is along decreasing $Y$-axis. (b) Gradual bending of an elastic beam of longer length $L=3$ cm (bending regime). The granular packing fraction is the same as in (a).}
  \label{FiberShape}
\end{figure}

\begin{figure}
\begin{center}
(a)\includegraphics[width=0.44\linewidth]{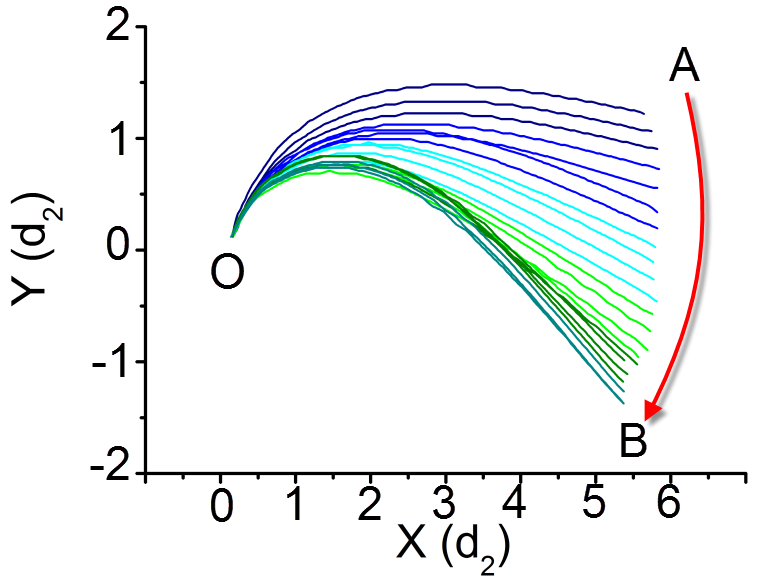}
(b)\includegraphics[width=0.44\linewidth]{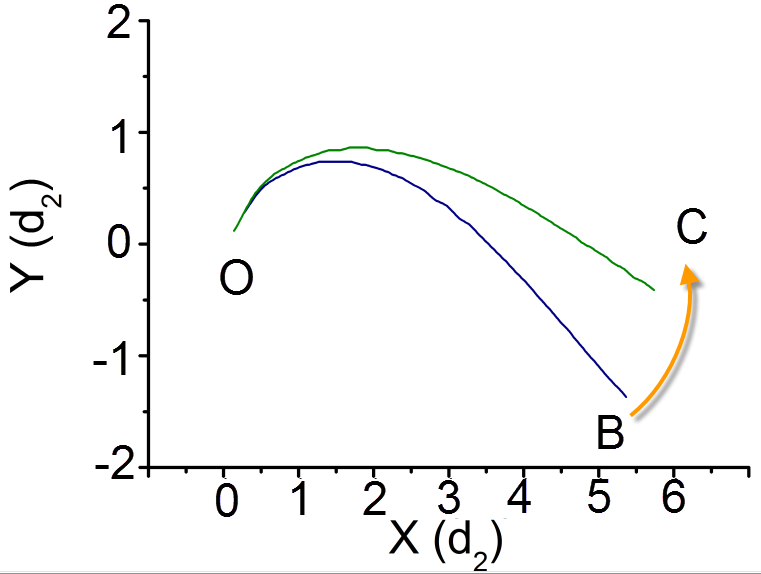}
\end{center}
\caption{Individual avalanche, loading and unloading of a fiber of length $L=3$ cm in a granular medium of packing fraction $\phi=80.94\%$ for a highly bent configuration (regime III).(a) Successive fiber shapes during a loading phase (plate displacement of $2U_0=d_2/6$ between each fiber shape from A to B) . (b) Unloading phase from B to C with a sudden elastic return of the fiber during one increment $U_0$ of plate displacement.} 
 \label{Avalanche}
\end{figure}

\begin{figure}
\begin{center}
(a)\includegraphics[width=0.9\linewidth]{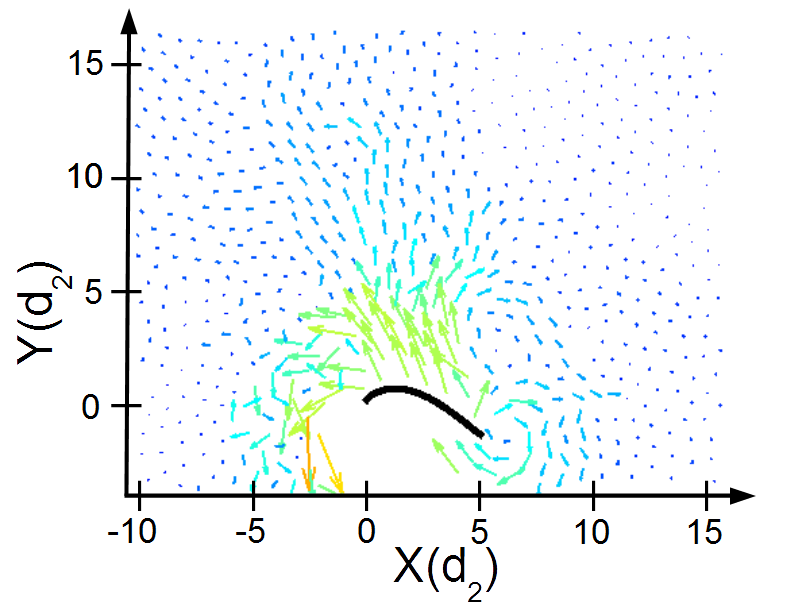}
(b)\includegraphics[width=0.9\linewidth]{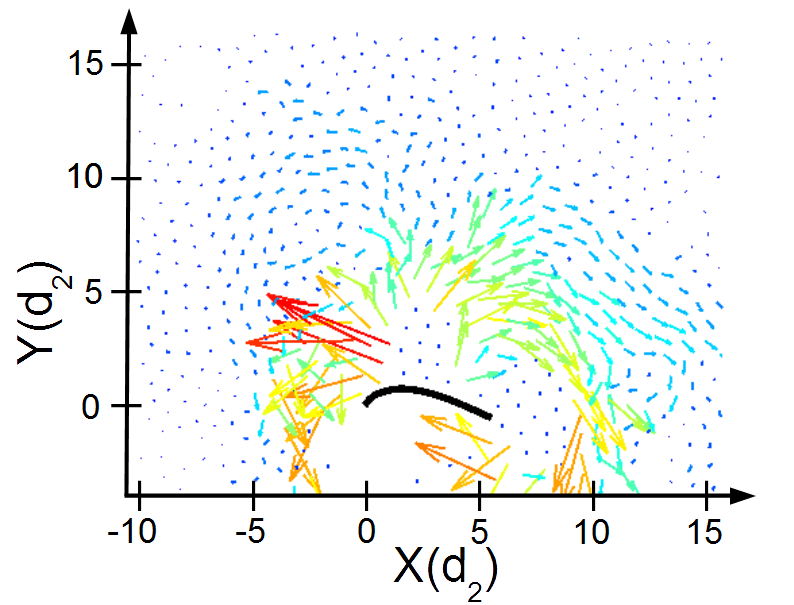}
\end{center}
\caption{ Map of the displacement field of grains (amplified by a factor of 20) in the plate frame between 2 consecutive images for the loading and for the unloading phases of Fig.~\ref{Avalanche}. The plate displacement $U_0$ between two consecutive images has been subtracted from each grain displacement.  (a) Between the two images just before the unloading event (b) during the avalanche BC.} 
 \label{AvalancheCarto}
\end{figure}

\section{Phenomenology: fluid-structure interaction in a granular flow\label{phenomenology}}

When immersed in a fluid, a flexible structure interacts with the
flow. In particular, depending on the rigidity of the structure and
the flow conditions, it can switch between two stable states: the
initial undeformed configuration and a deformed configuration that
reduces the drag force. Alternatively, for a given bending rigidity, there exists a critical length of the structure placed across the flow above which reconfiguration occurs.
  
A very similar phenomenology can be observed in the present experiment
which can be regarded as an interaction between a flexible beam and a
granular flow. In Fig.~\ref{FiberShape}a we show a series of
configurations of a short (rigid) fiber ($L=2$~cm) facing the granular
flow. Small fluctuations are observed around the stable straight
configuration, as if the fiber was jiggling around its initial position. The slight deflections are here caused by collisions with incoming grains.

In Fig.~\ref{FiberShape}b we show the contrasting case of a slightly longer fiber ($L=3$~cm) for the same granular packing fraction as before. In this case, the fiber experiences reconfiguration due to the flow and enters into a regime of bending. 

When the fiber has been highly tilted, fluctuations of the fiber positions are again observed but of a very different nature: in Fig.~\ref{Avalanche}a we show the
gradual bending of the deformed fiber due to an accumulation of
incoming grains and in Fig.~\ref{Avalanche}b its sudden
partial recovery. In Fig.~\ref{AvalancheCarto}, we represent the corresponding grain displacements in the plate frame (by subtracting the plate displacement $U_0$ between 2 consecutive images) as if the fiber was penetrating the granular material. Fig.~\ref{AvalancheCarto}a shows the displacement field between the two images just before the unloading event. It is characterized by longe range displacements in front of the fiber and small recirculations on both sides. Fig.~\ref{AvalancheCarto}b shows the displacement field corresponding to the unloading phase, with the fiber returning from position B to position C. In contrast with the loading phase, the elastic return of the fiber is associated to huge recirculations on both sides. Therefore the sudden
change of the fiber conformation can be associated to a large
reconfiguration of the grains, in other words a burst of the velocity
field of the granular flow. Such an alternation of smooth and regular loading stages interspersed
with sudden unloading events is typical of granular
avalanches and can also be associated to the turbulent-like velocity fluctuations
reported in dense granular flows~\cite{Radjai-PRL02,Atman-AIP13,Mizuno-SM16}.

\begin{figure}
\includegraphics[width=0.95\linewidth]{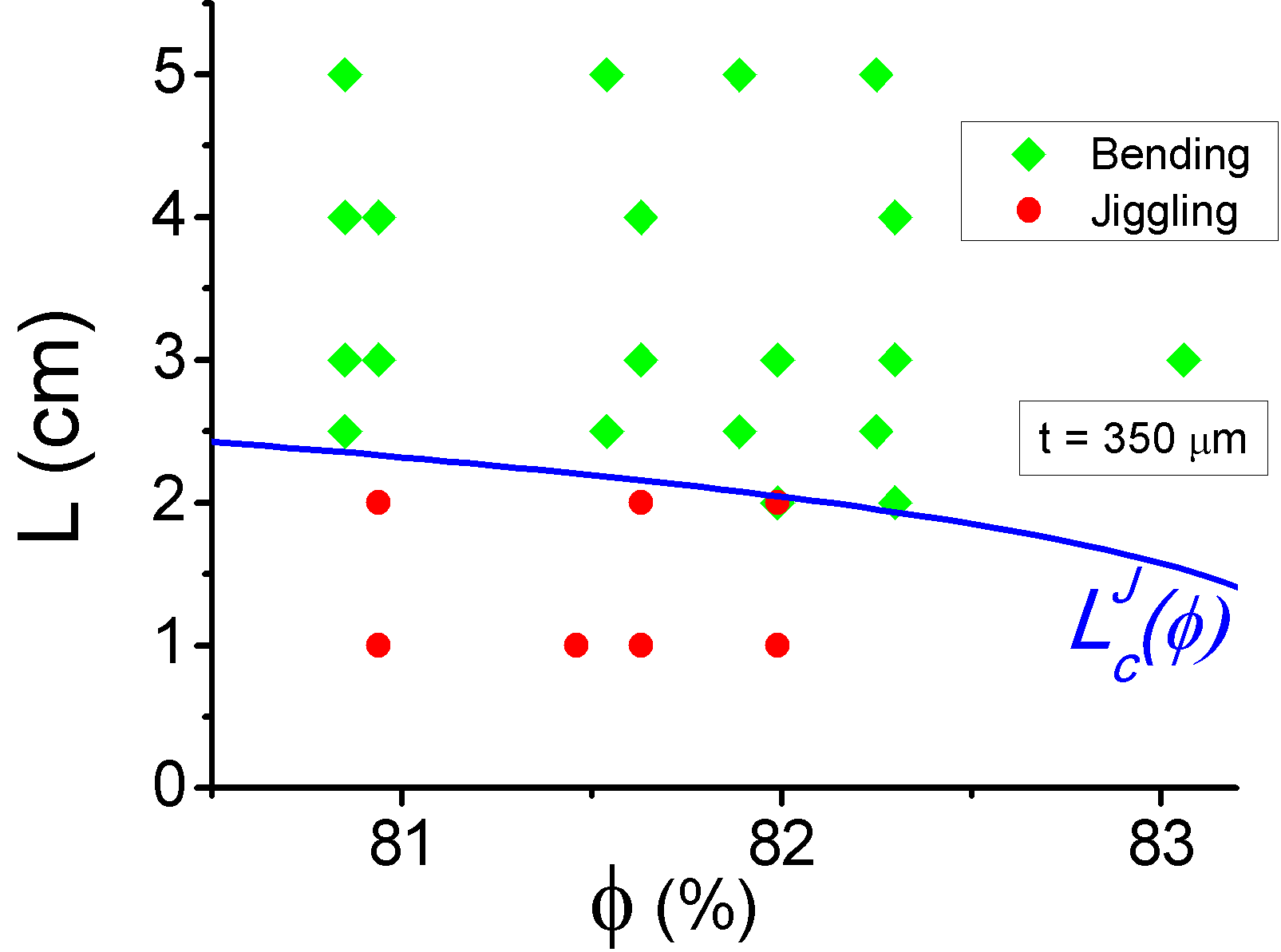}
 \caption{Phase diagram in the plane $L-\phi$. Experiments vs prediction. The symbols represent the experiments performed by varying the fiber length $L$ or granular packing fraction $\phi$ with green diamonds for the bending regime and red circles for the jiggling regime. The blue curve is the  predicted elasto-granular length from eq.(\ref{Critical-Length-jammed}) that separates the jiggling from the bending regimes.}
  \label{phase-diagram}
\end{figure}

In Fig.~\ref{phase-diagram} we summarize our observations of the straight (jiggling regime)
or deformed conformation (bending regime) of the fiber when varying its length $L$ as well as the
packing fraction $\phi$ of the granular medium.
  
In the present experiment, the flexible structure is initially facing the flow in a geometry of penetration. Therefore one would naturally introduce the critical force for fiber's buckling. In the case of a non embedded fiber, this force is given by the Euler force $F_{C}$ with clamped-free extremities. It drastically depends on the fiber length $L$ as:

\begin{equation}
F_{C}(L)=\frac{\pi^2}{4}\times\frac{EI}{L^2}
\label{eqn1}
\end{equation}

\noindent
where $I$ is the quadratic moment of the fiber corresponding to a non
embedded fiber buckling in the $(X,Y)$ plane. Taking into account the
rectangular cross section of the fiber yields
$I=\frac{h\:t^3}{12}$. Thus varying $L$ is a way to vary the maximum
acceptable loading force (or equivalently the maximum penetrative
force) in a case of a pure axial loading. For example, the typical
bending rigidity for a beam of length $L$=3~cm and thickness
$t$=350~$\mu$m is $EI=4.07\:10^{-5}~N.m^2$ and the corresponding Euler
force would be $F_{C}(L)\approx$~0.11~N.
For comparison, the bending force produced by the fiber tip displacement on one grain size $d$ (the typical size of obstacle encountered by the fiber) is:

\begin{equation}
F_{B}(L)=\frac{3EId}{L^3}
\label{eqn2}
\end{equation}

Thus the bending force will be $F_{B}(L)\approx 0.02$~N for $L$=3~cm, which is much smaller than the corresponding buckling force for the same geometrical parameters of the fiber. In any cases, the thickness $t$ of the fiber is small compared with the diameter $d$ of a grain with ${t}/{d}\approx0.08$, such that the probability of a force purely acting along the axis of the beam will be negligible. Therefore the main mechanism producing reconfiguration of the fiber will be attributed to bending and not buckling, contrary to the recent experiment of ~\cite{Holmes-EML16}.

In studies of interactions between fluids and flexible structures it is
customary to design a non dimensional number that compares drag and
elastic restoring forces in order to determine the stability of the
flexible object~\cite{Lindner-ARFM17,Gosselin-JFM10}. In the present context, a first
naive scaling analysis may consist in building such a Cauchy-like
number from the comparison between the friction force associated to the contact between the bottom plate and an
individual grain $f_n = \mu \rho g \pi d^2 h/4$ and the bending force
associated to a deflection $\delta$ of the tip of the fiber equivalent to one grain size $d$, i.e. $f_b= 3EId/L^3$:
\begin{equation}
c_Y^G = \frac{f_n}{f_b}= \mu  \frac{\pi\rho g d}{E}\left(\frac{L}{t}\right)^3
\label{Cauchy-individual}
\end{equation}
In this analysis that  corresponds to a very dilute gas-like regime where the flexible beam only experiences collisions with isolated grains, the transition between straight and deformed configurations of the beam would occur at $c_Y \approx 1$ for a critical length:
\begin{equation}
L_c^G \approx t \left(\frac{E}{\mu\pi\rho g d} \right)^{1/3}
\label{Critical-Length-individual}
\end{equation}
With the experimental parameters used in the present set-up, we get
$L_c^G \approx 7.5$~cm, i.e. below this length scale the flexible
intruder should not experience bending reconfiguration.

However, as shown in Figs.~\ref{FiberShape}-\ref{Avalanche}, fully bent
configurations were actually observed for much shorter fibers
(L~=~3~cm) than $L_c^G$. This diluted regime argument thus underestimates the drag
force exerted by the granular flow. In order to account for the effect
of a denser liquid-like flow, one can consider the friction force associated to
the grains present in the area that a slightly bent beam should sweep over
to recover its initial straight shape. For a tip deflection of $\delta$ the swept area $\cal{A}_\delta$ is of the order of ${\cal A}_{\delta} \approx {\delta L}/{2}$  and the corresponding number of grains that are pushed back is ${\phi \cal{A}_{\delta}}/{s}$ where $s$ is the surface occupied by one grain on the bottom plate. The associated friction force will be $F_n= \mu  {\phi\rho g h \delta L}/{2}$. Comparing this friction force $F_n$ to the elastic restoring force $F_b={8 E I \delta}/{L^3}$ for an homogeneous and perpendicular loading  on the whole fiber leads to:  
\begin{equation}
c_Y^L = \frac{F_n}{F_b}= \mu  \frac{3\phi\rho g t}{4E}\left(\frac{L}{t}\right)^4 \;,
\label{Cauchy-dense}
\end{equation}
and 
\begin{equation}
L_c^L \approx t \left(\frac{4E}{3\mu\phi\rho g t} \right)^{1/4} \;
\label{Critical-Length-dense}
\end{equation}
which gives now a slightly lower value of the critical length,
$L_c^L \approx 5.5$~cm, but which remains much larger than the
experimental observations despite the change of scaling exponent. This
means in particular that the friction forces exerted by the grains are
underestimated in the present argument.

Close to the jamming packing fraction, interactions of the flexible beam with
the flow are actually expected to strongly alter the structure of the
granular medium~\cite{Rognon-EPL08}. Transient jammed clusters can
thus form in the vicinity of a wall~\cite{Rognon-JFM15} or a rigid
intruder~\cite{Kolb-PRE13,Kolb-GM14}, and more generally by the
application of shear stress~\cite{Behringer-Nat11}. 

In the present context, the elastic return of the deflected fiber to
its straight conformation not only induces the displacement of the
grains present in the swept over area but also the building of a
jammed cluster. 
The area of the latter can be estimated by a simple conservation argument. Assume an initial grain packing fraction $\phi$, the
area $\cal{A}_J$ of the cluster that reaches the jamming packing fraction
$\phi_J$ after it has absorbed the grains swept over by the elastic
fiber is:
\begin{equation}
{\cal A}_J\left(\phi_J-\phi \right) = \frac{\delta L}{2} \phi \;.
\end{equation}
In this close-to-jamming regime, we can thus build a Cauchy number
with a drag force that results from the friction force of the jammed
cluster $F_J=\mu\rho g h \phi_J\cal{A}_J$. Here, the vicinity of the jamming
density limit thus simply induces an amplification of the friction
force by a factor $\phi_J/(\phi_J-\phi)$ so that we get for the Cauchy
number:
\begin{equation}
c_Y^J = \frac{\phi_J}{\phi_J - \phi} c_Y^L  = \mu   \frac{\phi\phi_J}{\phi_J - \phi}\frac{3\rho g t}{4E}\left(\frac{L}{t}\right)^4 \;,
\end{equation}
and for the critical length:
\begin{equation}
L_c^J \approx t \left(\frac{\phi_J - \phi}{\phi\phi_J}\frac{4E}{3\mu\rho g t} \right)^{1/4} \;.
\label{Critical-Length-jammed}
\end{equation}
This critical length can be viewed as an elasto-granular length as also defined in~\cite{Holmes-arxiv17} in analogy with other works on elasto-capillary phenomena \cite{Roman-JPCM10}.

The stability of the flexible beam thus appears to be controlled by two parameters: the rigidity (here the length) and the packing fraction of the granular medium.
Using the value $\phi_J = 0.8356$ determined in
Ref.~\cite{Kolb-PRE13}, this leads, for an initial density
$\phi=0.815$ to $L_c^J \approx 2.5$~cm, a value that is more
consistent with our experimental observations. In the phase diagram of
Fig.~\ref{phase-diagram} we see that the expression of the elasto-granular length $L_c^J(\phi)=L_c$ from eq.~(\ref{Critical-Length-jammed}) gives indeed a reasonable account of the transition between the straight or deformed conformation of the fiber.

\section{Development of the transition\label{transition}}

In contrast to buckling instability, the switching of the flexible
beam from its initial straight conformation to a bent one is gradual
and requires reorganization of the granular medium. We focus here on
the development of this transition.

\begin{figure}
\vspace{5pt}
\includegraphics[width=0.7\linewidth]{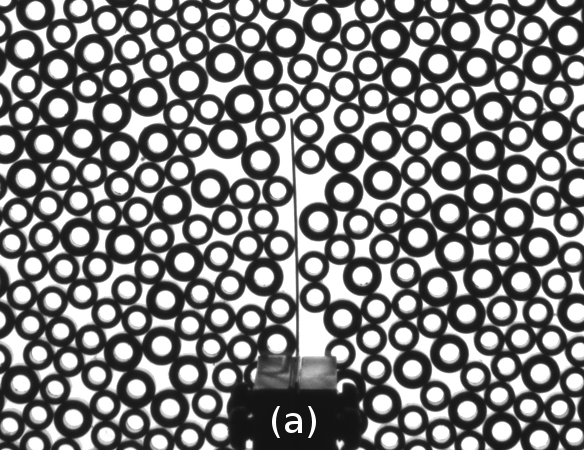}	
\vspace{5pt}\\
\includegraphics[width=0.7\linewidth]{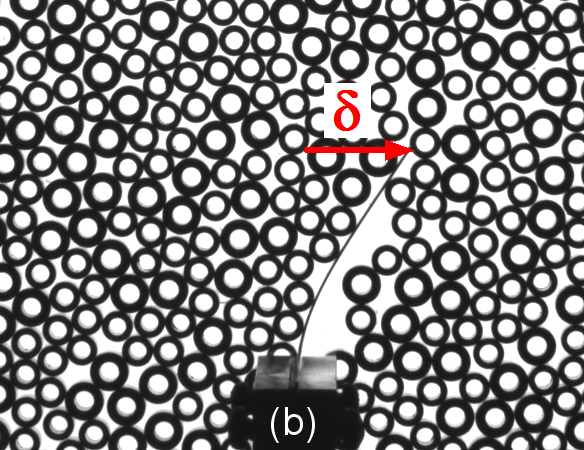}	
\vspace{5pt}\\
\includegraphics[width=0.7\linewidth]{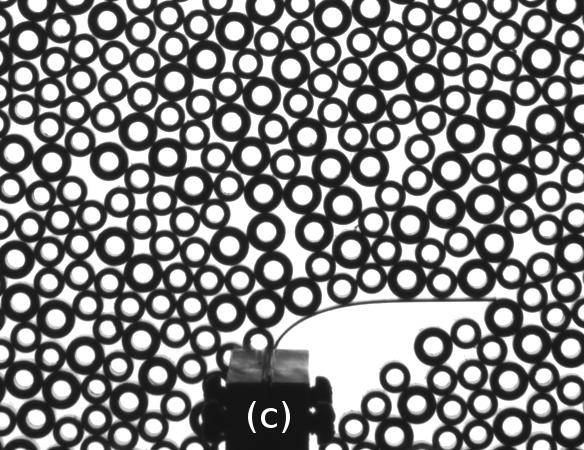}	
\vspace{7pt}\\
\includegraphics[width=0.85\linewidth]{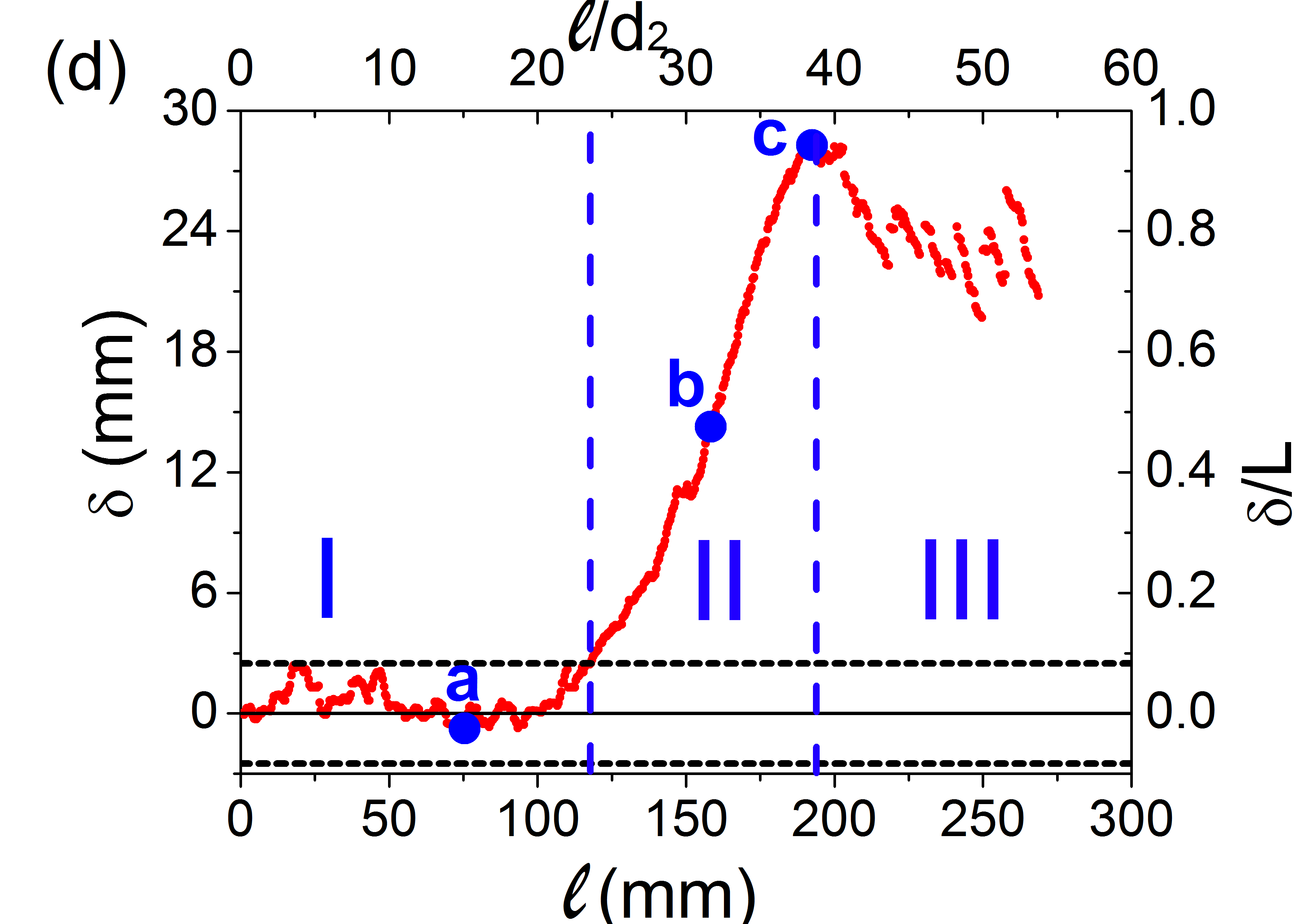}
  \caption{Irreversible deflection of the bending fiber of length $L=3$ cm in a granular medium of packing fraction $\phi=80.94\%$ for different penetration distances $\ell$. (a) $\ell$~=~15.1~$d_2$ (b)   $\ell$~=~31.7~$d_2$  (c) $\ell$~=~38.5~$d_2$  (d) Evolution of the lateral deflection $\delta$ of the free fiber extremity as a function of the penetration distance $\ell$. The horizontal dotted lines represent the maximum excursion of $\delta$ during regime I. The three successive regimes I, II and III are separated by the vertical blue dotted lines.}
  \label{ImaRegimes}
\end{figure}

A typical sequence (zoomed on the fiber) is shown in Fig.~\ref{ImaRegimes} for an intermediate fiber length $L=3$~cm and a granular packing fraction $\phi=80.94 \%$. The initial location of the fiber is along the vertical of the figure ($Y$-axis). In a first stage, the fiber experiences only small lateral deflections and fluctuates around its straight initial position. For example, in Fig.~\ref{ImaRegimes}a, the fiber is slightly deflected on the left side and the lateral displacement of the free extremity is almost not visible. The granular material is still homogeneous, while grains are contacting the fiber on both sides.  After a given traveling distance, the fiber bends irreversibly on one side with no possibility of returning back to its straight shape. While progressing inside the granular material, the fiber continues to bend (Fig.~\ref{ImaRegimes}b) with a clear lateral deflection on the right side, indicated by the horizontal (red) arrow along the $X$-axis. Associated to this deflection, we clearly observe the appearance of a cavity downstream of the fiber. As the whole granular material still flows along the $Y-$axis, the fiber continues to bend and the free extremity might reach its maximum lateral deflection (Fig.~\ref{ImaRegimes}c) when most of the fiber axis is placed perpendicularly to the initial flow. At this stage, the area of the cavity is larger. In the following we will focus on this bifurcation in the penetration behavior by introducing some quantitative descriptors.

\subsection{Deflection of the fiber tip}

One way to quantitatively characterize the penetration of the fiber
relative to the granular medium is to compute the lateral deflection
$\delta$ (along the $X$ axis) of the free extremity of the fiber as a
function of the distance $\ell$ traveled by the plate (or equivalently
the displacement of the fiber anchoring point relative to the granular
medium). The evolution of $\delta$ as a function of $\ell$ associated
to the experiment shown in Figs.~\ref{ImaRegimes}abc is shown in
Fig.~\ref{ImaRegimes}d.

\begin{figure}
 (a) \includegraphics  [width=0.95\linewidth] {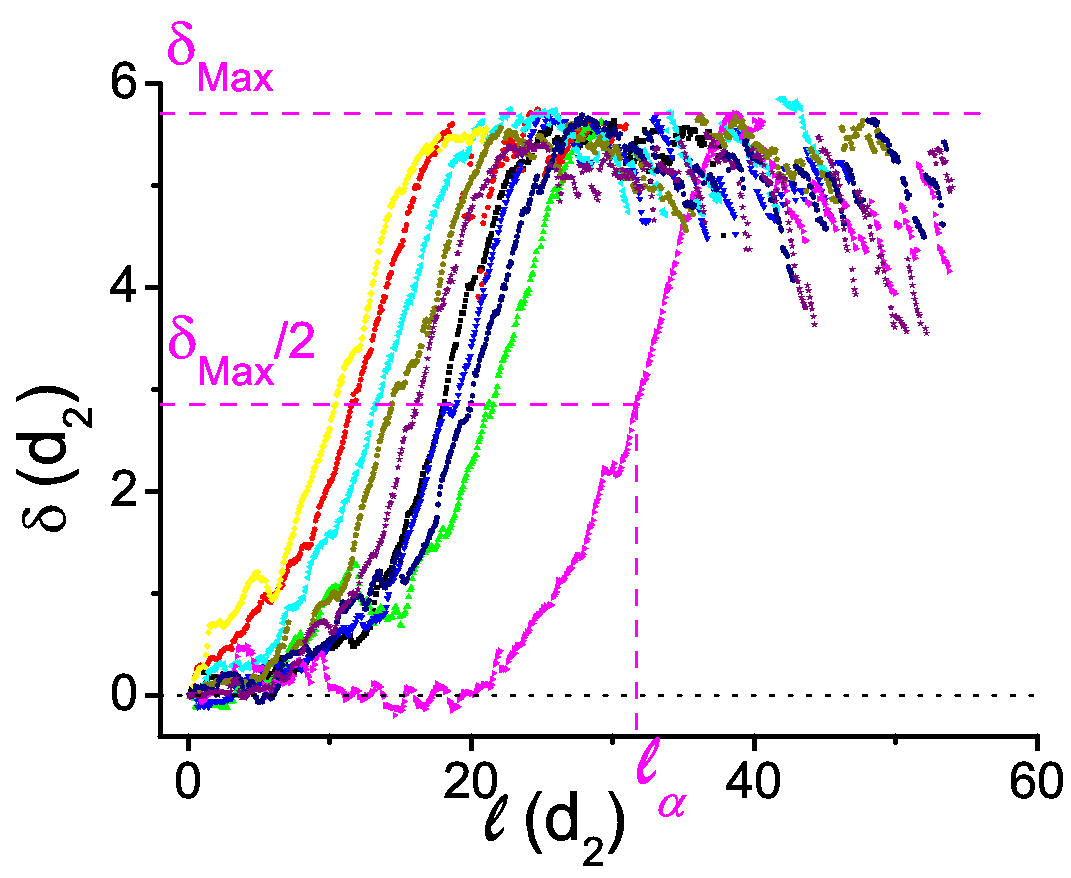}
 (b) \includegraphics  [width=0.95\linewidth] {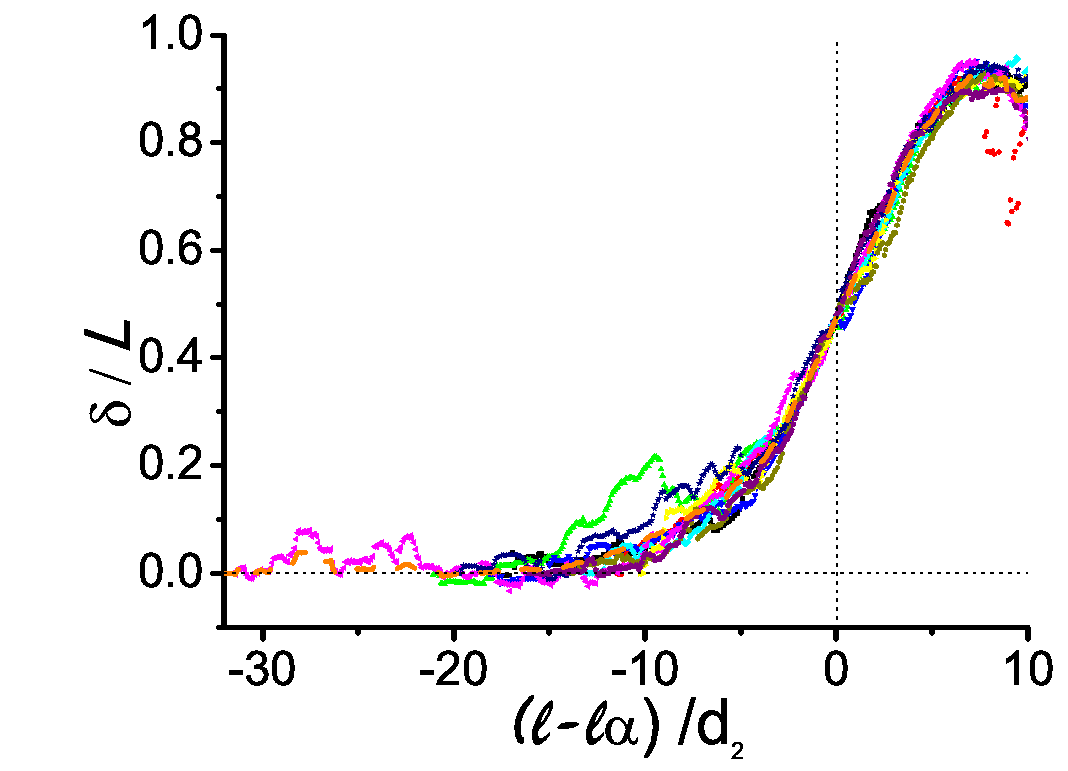}
  \caption{(a) Lateral deflection $\delta$ of the free extremity of the fiber as a function of the plate displacement (or penetration distance) $\ell$ for 10 independent experiments performed with the same fiber length $L$=3~cm and granular packing fraction $\phi$=80.94\%.  Both quantities are given in diameters of a large grain $d_2$. All the 10 curves reach the same maximal value $\delta_{Max}$, with $\ell_{\alpha}$ the penetration distance obtained for $\delta_{Max}/2$.(b) Same curves as (a) expressed as a function of the resetted penetration distance $\ell-\ell_{\alpha}$.}
  \label{Deflection}
\end{figure}

In order to give a more systematic account of this phenomenon, we
report in Fig.~\ref{Deflection} the evolution of the deflection $\delta$
as a function of the penetration distance $\ell$ expressed in diameters
of a large grain ($d_2$) for ten independent experiments performed for
a fiber length $L=3$~cm and a granular packing fraction
$\phi=80.94\%$. At the end of an experiment, the fiber is observed to
bend either toward the right (along $X>0$) or toward the left side
(along $X<0$). In order to compare experiments, the $\delta$ values
corresponding to experiments where the fiber eventually bends toward
the left side have been inverted in Fig.~\ref{Deflection}, such that
the final value of $\delta$ is always positive.

From Fig.~\ref{ImaRegimes}, Fig.~\ref{Deflection} and from the corresponding experimental images, we recover the three stages identified above:

- In a first regime  (I), the lateral deflection $\delta$ is observed to fluctuate around zero. The fiber continuously penetrates the granular medium and keeps in average its initially straight shape. When the fiber is contacting grains at its tip, it is slightly deflected on one side or the other but the maximal amplitude of $\delta$ does not exceed the radius of a grain. Thus the extremity of the fiber experiences an erratic and fluctuating motion, depending on the local arrangement of the grains encountered by the fiber. There is only few interaction between the fiber and individual grains, and the wake left behind the fiber is reduced to a small elongated cavity almost not visible.

- After a penetration distance $\ell_{1}$ (indicated by the vertical blue dotted line in Fig.~\ref{ImaRegimes}d) that strongly depends on the
realization, the fiber irreversibly bends towards one side (left or
right). We checked the absence of bias in the direction of
deflection. This regime II is characterized by a progressive and
continuous bending of the fiber, shown by the regular increase of
$\delta$ with the pentration distance $\ell$.  This increase appears to
be quite similar for the different experiments of
Fig.~\ref{Deflection}.

- A third regime appears when the extremity of the fiber roughly lies
perpendicular to the direction of the plate translation $X$, ie
$\theta(s=1)\approx \pi/2$. Then the corresponding lateral deflection
reaches its maximal value $\delta_{Max}$. For the examples of Fig.~\ref{Deflection}, the fibre then fluctuates
around this new bended configuration. In the case of a pinned
rigid fiber, the maximum possible value of the deflection would be
$\delta_{Max}$=$L$. In our case, the anchoring point is clamped and
the fiber is flexible. Then the observed $\delta_{Max}$ is slightly
smaller than $L$, but has a constant value independent of $\phi$. For
example, the average experimental values are $\delta_{Max}$=
(0.952$\:\pm\:$0.005)$L$ for $L$=3~cm.

\begin{figure}
 (a) \includegraphics  [width=0.95\linewidth] {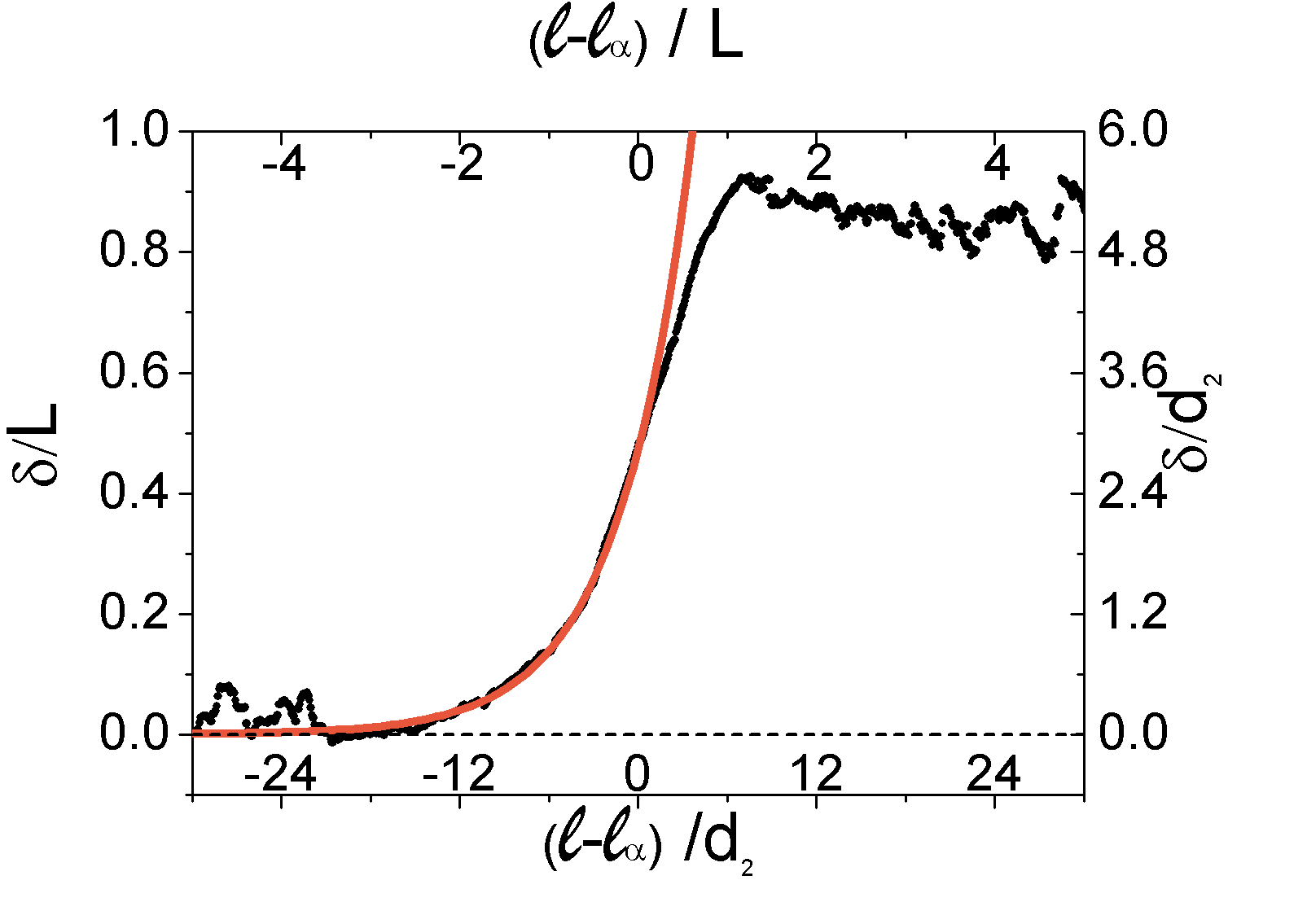}
 (b) \includegraphics  [width=0.95\linewidth] {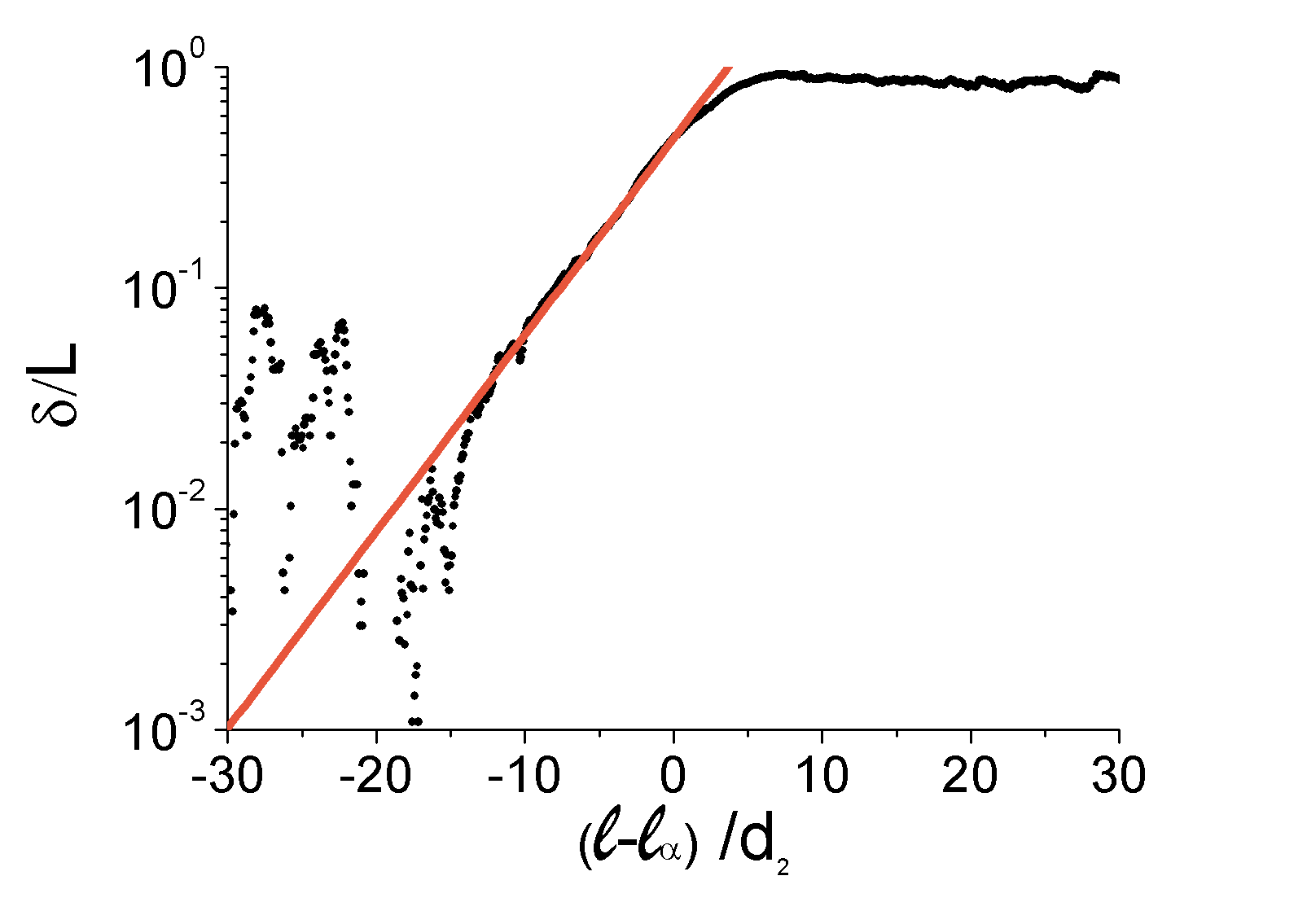}
  \caption{Average of the lateral deflection $\delta$ of the free extremity of the fiber over the ten experiments of Fig.~\ref{Deflection} as a function of the rescaled plate displacement $\ell-\ell_{\alpha}$. Both quantities are normalized by the diameter of a large grain $d_2$ (bottom and right axes) or by the fiber length (left and top axes).The orange curve is an exponential fit of the bending. (a) normal scales (b) semi-log scales.}
  \label{DeltaMoy}
\end{figure}

As it appears clearly in Fig.~\ref{Deflection}a, we observe two
fluctuating regimes separated by a transition that in contrast appears
to be very deterministic. This deterministic aspect appears even more
clearly in Fig.~\ref{Deflection}b after a resetting of the abscissa.
Starting from the observation that the maximal deflection
$\delta_{Max}$ only depends on the length of the fiber we could indeed
locate the traveling distance $\ell_{\alpha}$ corresponding to the
deflection value $\frac{\delta_{Max}}{2}$, the half of the average
maximal deflection for each experiment $\alpha$. Then we could plot in
Fig.~\ref{Deflection}b the deflection $\delta$ as a function of the
resetted penetration distance distance $\ell-\ell_{\alpha}$. The curves collapse rather well such that the average over the ten shifted experiments can be computed (Fig.~\ref{DeltaMoy}a). 
Interestingly the first part of the bending regime can be fitted by an exponential growth (orange curve superimposed on the averaged experimental deflection in Fig.~\ref{DeltaMoy}a). The same plot is presented in semi-log scales in Fig.~\ref{DeltaMoy}b.
This fit provides a characteristic length $\lambda$ for the kinetics of bending according to the formula $\delta=\delta_0 exp(\ell / \lambda)$. For the example of Fig.~\ref{DeltaMoy}, $\lambda$=(4.89$\:\pm\:$0.04)$d_2$. The length $\lambda$ has been obtained for all batches of experiments where bending occurred. Its value normalized by the $L_c$ value from eq.~(\ref{Critical-Length-jammed}) associated to the packing fraction $\phi$ is plotted in Fig.~\ref{LambdaLc} as a function of the rescaled fiber length, that is $\epsilon=\frac{L-L_c}{L_c}$. 
The $\epsilon$ value can be viewed as an order parameter: For $\epsilon<0$, no irreversible bending occurs, the fiber stays in the jiggling regime, while for $\epsilon<0$, bending occurs.

\begin{figure}
\begin{center}
  \includegraphics[width=0.75\linewidth]{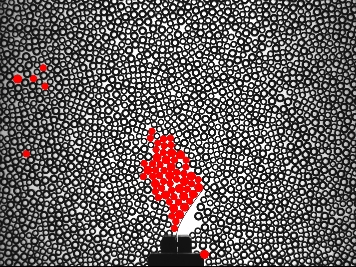}\\
  \vspace{4pt}
\includegraphics[width=0.75\linewidth]{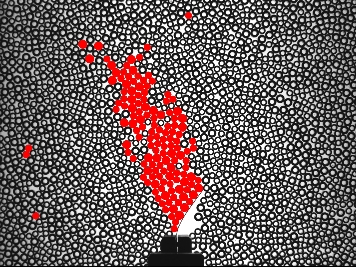}\\	
  \vspace{4pt}
\includegraphics[width=0.75\linewidth]{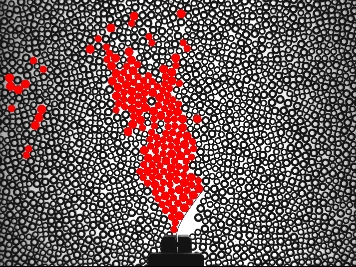}	
\end{center}  
\caption{Growth of a ``cluster'' of slow grains (red-labeled) for three successive images starting from a penetration distance $\ell$=24.1$d_2$ for a granular packing at $\phi=80.94\%$ and $L$~=~3~cm.}
\label{cluster-growth}
\end{figure}





\subsection{Asymmetry of the packing fraction}

As clearly visible in Fig.~\ref{ImaRegimes}, when the fiber enters the
deflected regime and that the lateral deflection is larger than a
grain size, we observe the formation of a cavity empty of grains
downstream of the fiber. A similar trend was observed in the framework
of the penetration of a rigid intruder in a granular
layer~\cite{Kolb-PRE13}.

Conversely, we expect an accumulation of grains, more specifically the
formation of a jammed cluster upstream of the fiber. Such an evolution
is indeed observable in Fig.~\ref{cluster-growth} where we present
three successive snapshots of an experiment of penetration of a fiber
of length $L=3$~cm in granular medium of packing fraction $\phi=80.94\%$.

\begin{figure}[t]
\begin{center}
(a)\includegraphics[width=0.44\linewidth]{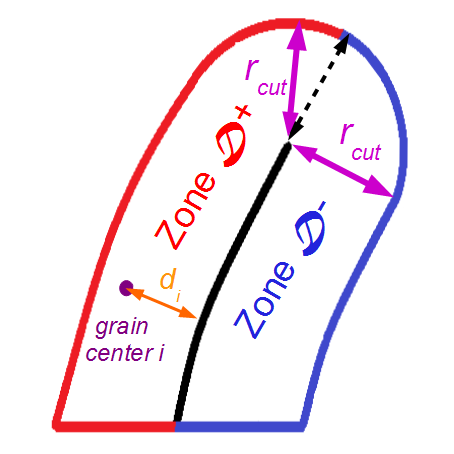}
(b)\includegraphics[width=0.44\linewidth]{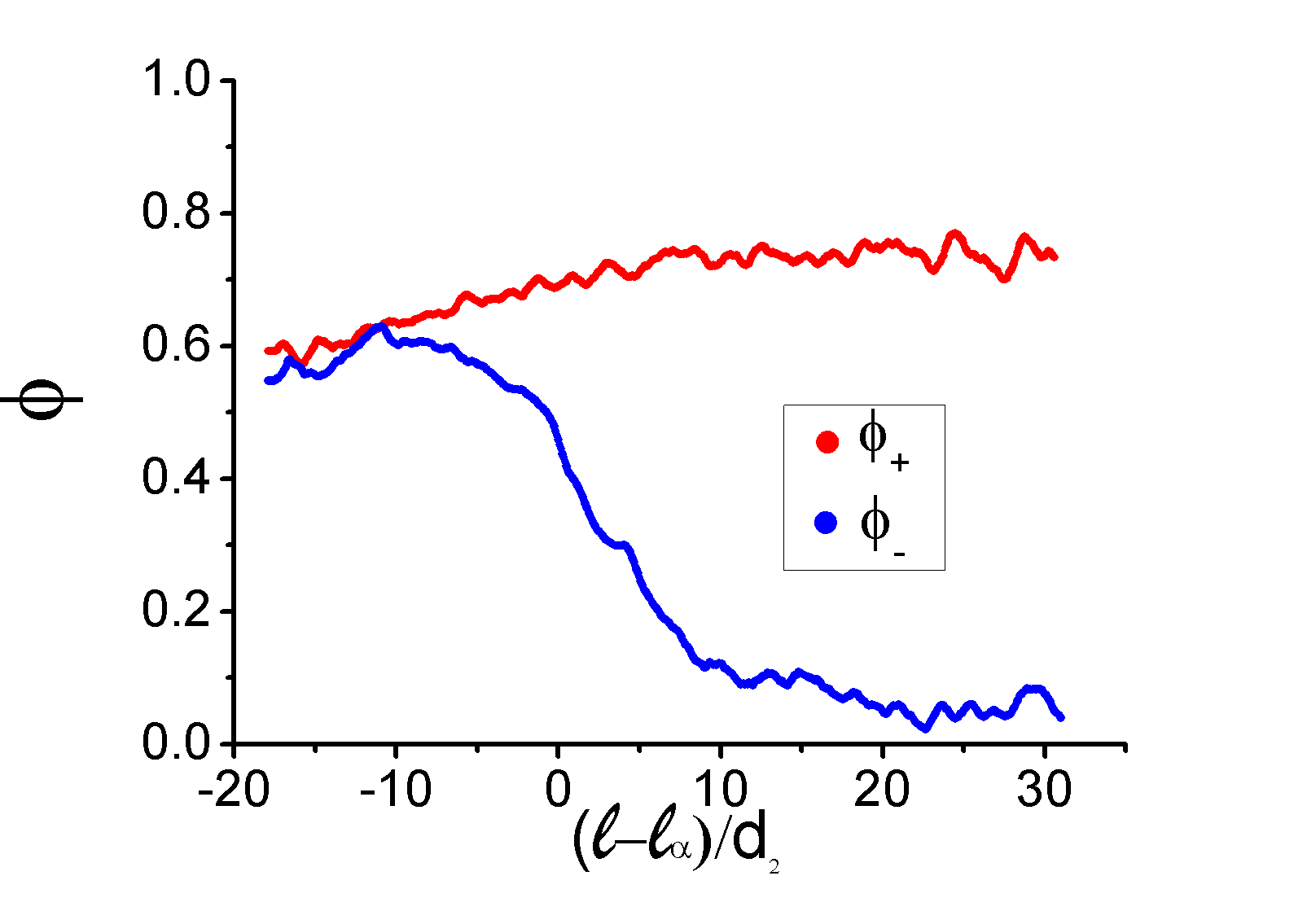}
(c)\includegraphics[width=0.44\linewidth]{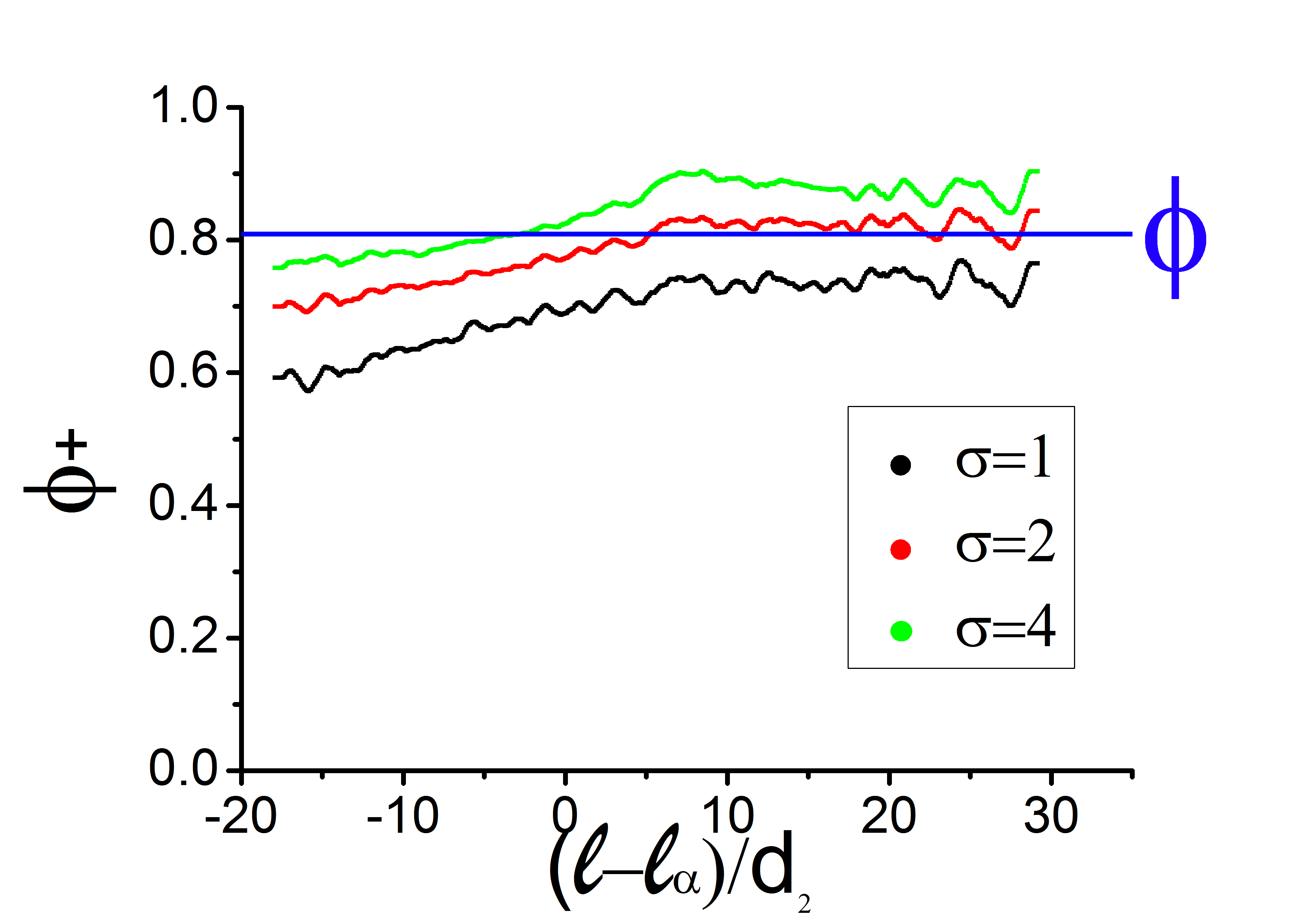}	
(d)\includegraphics[width=0.44\linewidth]{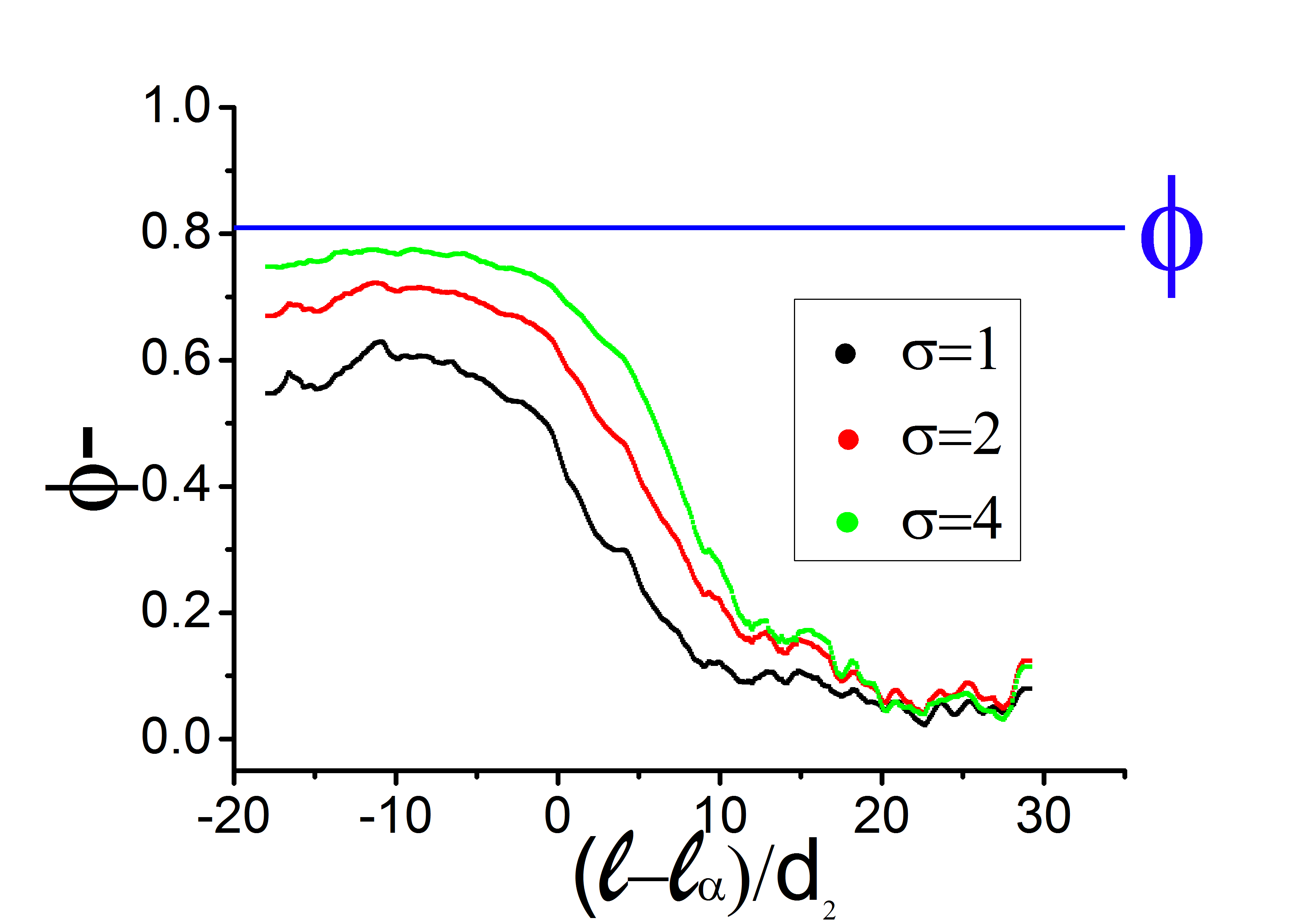}	
\end{center}
\caption{Development of  a packing fraction contrast between the two sides of
    the bending fiber. (a) Schematic diagram of the zones upstream and downstream the fiber for calculation of $\phi_+$ and $\phi_-$. (b) Lateral packing fractions $\phi_+$ and $\phi_-$ (sliding window averaged over 10 points and over 10 experiments) as a function of the rescaled penetration distance $\ell-\ell_{\alpha}$ normalized by $d_2$ for a fiber length $L=3$~cm in a granular medium with a macroscopic packing fraction $\phi=80.94\%$ (indicated by the blue horizontal line in (c) and (d)): Effect of the choice of coarse-grained length scale $\sigma$ on the calculation of (c) $\phi_+$ and (d) $\phi_-$.}
  \label{density-contrast}
\end{figure}

In the present experiment, we recall that the anchoring point of the flexible
intruder is fixed in the laboratory frame while the granular medium
lies on a glass plate that is displaced at a constant velocity with $U_0$, the typical displacement of the plate between 2 successive images. The
grains that accumulate upstream of the fiber are thus expected to be
slowed down with respect to the mean flow. In
Fig.~\ref{cluster-growth}, we show the distribution of the slowest
(red-labeled) grains, here selected on a simple threshold displacement
criterion: We labeled the grains $i$ whose amplitudes of displacement between 2 successive images in the laboratory frame are $u_i < u_T= 0.048 U_0$. As expected, we observe a
strong clusterization of these slow grains upstream of the flexible
beam. Although only qualitative, this observation supports the idea
of the development of a jammed cluster of grains ahead of the fiber.

The penetration of the flexible beam in the granular medium thus
leads to a clear symmetry breaking in the local packing fraction.  In
order to quantify this effect, we computed the coarse-grained packing
fraction in the vicinity of the fiber.


Here we adapted the computation of the coarse-grained packing
fraction~\cite{Goldenberg-EPL07} to define over a coarse-grained
length scale $\sigma$ a left and a right packing fraction of grains on either
side of the fiber. As sketched in
Fig.~\ref{density-contrast}a, we define two domains ${\cal D}^+$ and
${\cal D}^-$ from either side of the fiber such that the distance of any
point of a domain to the fiber remains below a cut-off value $r_{cut}$.
The two packing fractions $\phi_+$ and $\phi_-$ are then computed as:

\begin{equation}
  \phi_\pm = \frac{1}{\alpha_\pm} \sum_{g_i \in {\cal D}^\pm}
 A_i \exp \left( -  \frac{d_i^2}{\sigma^2} \right)\;,
  \label{lateral-density}
  \end{equation}

\noindent
where $A_i$ is the area occupied by the grain $g_i$ and $d_i$ is its distance
to the fiber. Here the parameter $\sigma$ is the coarse-graining
length and we use $r_{cut}=4\sigma$. The normalization factor
$\alpha_\pm$ is obtained by the expression:
\begin{equation}
\alpha_\pm = \int_{\vec{r}\in {\cal D}^\pm} \exp \left( -  \frac{d(\vec{r})^2}{\sigma^2} \right) dxdy \;,
\label{lateral-density-normalisation}
\end{equation}
where $d(\vec{r})$ is the distance to the fiber of a point $\vec{r}$.

In Fig.~\ref{density-contrast}b, we show the evolution of the lateral
packing fractions $\phi^+$ and $\phi^-$ obtained for a coarse-graining length
$\sigma=d_2$ after resetting the
abscissa as in Fig.~\ref{Deflection} by using the distance $\ell_\alpha$ and then averaging over 10 experiments. 
The bending transition appears very
clearly to be accompanied by i) a strong decrease of the downstream
packing fraction $\phi^-$ down to values approaching zero and ii) a slight
increase of the upstream $\phi^+$ up to a plateau value about
$\phi^+_{max}\approx 0.7$. 

Note that this value is far lower than the jamming packing fraction of the
present granular medium $\phi_J\approx 0.8356$ estimated in
Ref.~\cite{Kolb-PRE13}. However, the value of the maximum packing fraction of a
granular cluster is known to be significantly altered (decreased) in
presence of a wall~\cite{Weeks-PRE09}. A clear illustration of this effect is given in
Figs.~\ref{density-contrast}c and \ref{density-contrast}d where we
show the respective evolution of the upstream and downstream packing fractions
$\phi^+$ and $\phi^-$ with increasing values of the coarse-graining
length $\sigma$. We see that the larger the coarse-graining length,
the larger the maximum upstream packing fraction $\phi^+$, which for
$\sigma=4d_2$ reaches a plateau value close to the expected jamming
packing fraction.

By comparing Fig.~\ref{density-contrast}b with Fig.~\ref{DeltaMoy}, we observe that the symmetry breaking in the packing appears when the fiber deflection typically exceeds the half diameter of a grain. These results thus give a quantitative support to our phenomenological
observations: the bending transition of the flexible beam is
associated to a structural transition of the flowing granular medium
with the formation of jammed cluster upstream of the fiber and the
development of a cavity downstream of the fiber.


\section{A simple model of bending induced compaction\label{model}}

\subsection{Bending induced compaction}

\begin{figure}
\includegraphics[height=6cm]{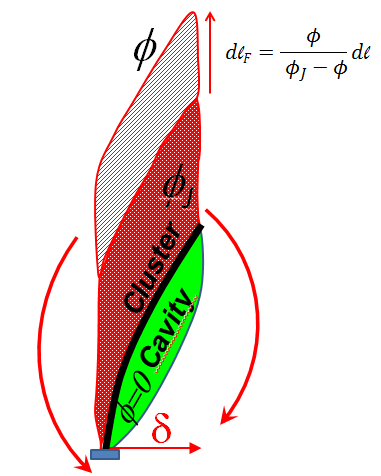}
 \caption{Schematic diagram of the formation of a cluster at the jamming packing fraction $\phi_J$ ahead of the fiber and a cavity empty of grains behind the fiber. Agregation of grains in front of the cluster through the advance of a compaction front and cluster erosion through recirculations of grains (red arrows) on both sides of the flexible intruder. The rest of the granular layer is at the packing fraction $\phi$.}
  \label{SchemaCluster}
\end{figure}

 As discussed in the previous section, the bending transition of the flexible beam is associated with an accumulation of grains upstream of
the fiber, leading to the growth of a cluster at the jamming density
$\phi_J$ as well as the development of a cavity empty of grains
downstream of the fiber (see the schematic diagram of Fig.~\ref{SchemaCluster}). 
The gradual bending of the flexible beam in the course of the
penetration of the granular medium tends to increase the section
of the intruder across the granular flow. We make the simple assumption that a larger intruder cross-section collects a larger number of grains in the cluster. Thus the number of grains $d{N_c}^+$ entering the cluster for an infinitesimal penetration distance of $d\ell$ is assumed to be proportional to the deflection $\delta$ along $X$. This accumulation of grains in the cluster, which acts as a solid block connected to the fiber, increases the force acting against the elastic restoring force on the fiber.

In the spirit of the compaction front observed in the work of Waitukaitis and al.~\cite{Waitukaitis-EPL13} and the three phase-model we developed for the penetration of a rigid intruder in a granular medium~\cite{Kolb-PRE13} we propose that the cluster grows faster for a packing fraction approaching the jamming transition. The compaction front (see Fig.~\ref{SchemaCluster}) delimits the interface between the cluster at $\phi_J$ and the rest of the granular layer at $\phi$ in front of the fiber. For each penetration step $d\ell$, the compaction front advances over a distance of $d\ell_F$ greater than $d\ell$. Indeed there is an amplification factor linked to the approach to the jamming transition: $\frac{\phi}{\phi_J-\phi}$, thus resulting in $d\ell_F$=$\frac{\phi}{\phi_J-\phi} d\ell$ for a one dimensional growth of the cluster as illustrated in Fig.~\ref{SchemaCluster}. Therefore the number of grains arriving in the cluster will evolve like: $\frac{d{N_c}^+}{d\ell}\:\propto\:\frac{\phi}{\phi_J-\phi}\delta$. 

However the compaction of a cluster of grains upstream of the fibre induces long range effects in the granular medium~\cite{Reddy-PRL11,Bouzid-EPJE15}. In particular, it is accompanied here by recirculation flows on both sides of the intruder that tend to erode the cluster and convect grains downstream of the fibre. These recirculations  are visible in the map of displacement fields of Fig.~\ref{AvalancheCarto} in the plate frame, when the displacement of the plate has been subtracted from each grain displacement. In our previous work on the penetration of rigid intruders \cite{Kolb-GM14}, it was observed that the lateral extent of recirculations was proportional to the diameter of the cylindrical intruder. Adapting this argument to the flexible intruder of spanwise length $\delta$ leads to a grain flux recirculating on both sides of the fiber which is proportional to $\delta$. The number of grains eroded from the cluster for each penetration step is thus expected to scale like $\frac{d{N_c}^-}{d\ell}\:\propto\:\delta$.
Hence the evolution of the number of grains $N_c$ in the cluster is the balance of two terms, an accretion term and an erosion term that both depend on $\delta$ with $\frac{d{N_c}}{d\ell}=\frac{d{N_c}^+}{d\ell}-\frac{d{N_c}^-}{d\ell}$.

We assume that the friction force resulting from contacts between grains and between grains and fiber
remains negligible with respect to the friction force resulting from the contact grains and the supporting plate. We also assume that during bending the accretion term ($A^+$) is larger than the erosion term ($A^-$), such that the cluster grows and induces an increment of the friction force $F_c$ with $\frac{dF_c}{d\ell}\:\propto\:\frac{dN_c}{d\ell}\:\propto\:\delta$.
During this transition regime we expect a quasi-static loading where the elastic force due to the bending $F_b$ compensates the friction force induced by the jammed cluster of grains upstream of the fiber. 
This leads to the following evolution law:

\begin{equation}
\frac{dF_b}{d\ell}=A^+\cdot \delta-A^-\cdot \delta = A\cdot \delta
\label{dF_b}
\end{equation}

Note that $A$ is a phenomenological parameter which has the dimension of a pressure.

Given the knowledge of the constitutive relation $F_b(\delta)$ that
relates the bending force of the flexible beam with its lateral
deflection, Eq.~(\ref{dF_b}) should thus give us an evolution law for the lateral deflection $\delta$ with respect to the
penetration of the fiber $\ell$.

In the limit of small deflection and for a uniform perpendicular force density along the fiber,
we can use the linear elasticity result $F_b = 8EI\delta/L^3$ and we get:

\begin{equation}
  \frac{d\delta}{d\ell} = \frac{ A\:L^3 }{8EI}\:\delta = \frac{\delta}{\lambda}
  \label{exponential-prediction}
\end{equation}  

We thus recover an exponential growth of the deflection of the
fiber with respect to the penetration. This was observed experimentally in Fig.~\ref{DeltaMoy} where $\delta=\delta_0 exp(\ell / \lambda)$ for the early stage of the bending. The identification of the exponential fit for the experimental values of $\delta$  with the evolution law of eq.~\ref{exponential-prediction} gives for the length $\lambda$ characteristics of the kinetics of bending:

\begin{equation}
  \lambda = \frac{ A\:L^3 }{8EI}
  \label{lambda}
\end{equation}

However the experimental evolution of the deflection with penetration eventually shows a
saturation which can not be reproduced in the framework of the linear
approximation of the elastic bending of the fiber, only valid for small deflection.

\begin{figure}
\includegraphics[width=0.98\linewidth]{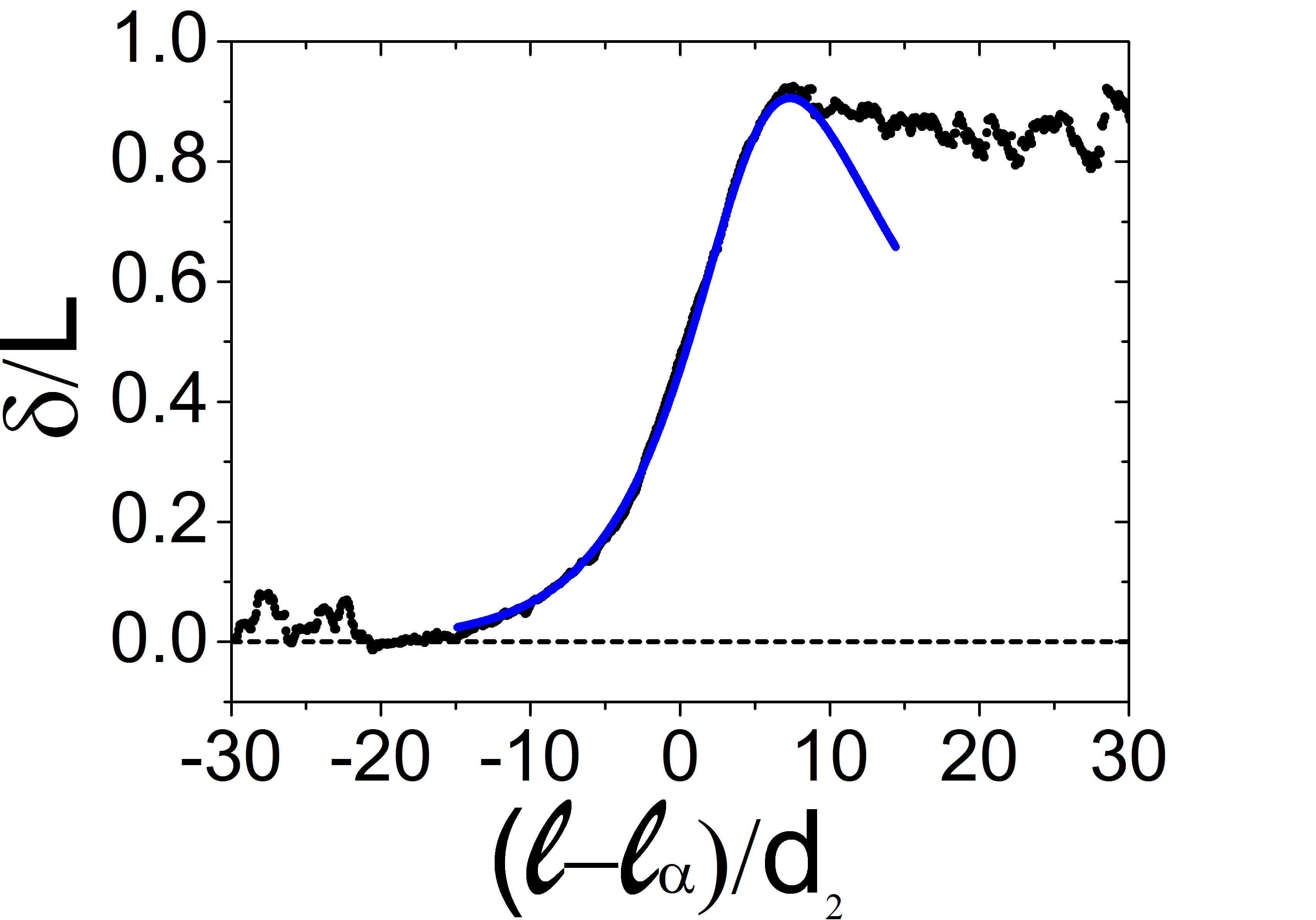}
 \caption{Bending-induced compaction model (blue curve) superimposed on the experimental evolution of the averaged lateral deflection (black curve) as a function of the rescaled penetration distance with the same experimental parameters as in Fig.~\ref{Deflection}). The best $A$ value obtained from the model is $A$=2199~Pa for this example.}
  \label{Elastica-compaction}
\end{figure}

\subsection{Elastica}

In order to account for the large deflection regime and for the saturation behavior, we set up a nonlinear numerical model of our experimental system. The fiber is modeled by a unidimensional inextensible centerline parameterized in a $2D$ space by the rotation angle along the arclength between the local tangent of the deformed and undeformed curve. Our model is geometrically exact in the sense that there is no restriction on the amount of deflection the inextensible fiber can take (the geometrical limit is that self-interactions are not allowed in our model). The equilibrium configuration of the weigthless fiber under an external density of force along its arclength is given by a $2D$'s Kirchhoff equation~\cite{Audoly-book2010}. To model the contacts of the grain on the fiber, we choose a density of forces that remains orthogonal to the fiber upon deformation (note that because of this non-conservative positional loading, the equilibrium equation cannot be derived from a mechanical energy formulation). Finally, we ensure a free boundary condition at one end and a torsional spring with a given stiffness on the other end to account for the possible imperfections of the experimental clamping.  

Using finite differences~\cite{ALazarus-JMPS13}, the second order partial differential equilibrium equation is discretized in a set of $N$ nonlinear algebraic equations with $N$ unknowns that are the rotational degrees of freedom of the discretized structure. Here, $N$ is the number of discrete elements modelling the fiber that has been set to $N=200$ to ensure numerical convergence. For a given discrete density of forces applied on the nodes of the discrete fibers, the equilibrium configuration is computed thanks to a classic Newton-Raphson algorithm. Note that the intensity of the density of forces has to be applied gradually from $0$ to the desired intensity to ensure proper convergence of the Newton-Raphson algorithm. This numerical process gives us the nonlinear relation $F_b(\delta)$ that we were seeking.

As shown by the blue curve in Fig.~\ref{Elastica-compaction}, this computation now
allows us to reasonably reproduce the full bending transition (regime II) of the fiber induced by the gradual compaction of grains upstream of the fiber. By tuning the value of $A$ to a given value, it is possible to reproduce the experimental evolution of deflection till its maximum value. The model does not allow to reproduce the behavior beyond (regime III), in particular the plateau observed for this fiber length $L$=3~cm, as the elastica calculations are based on the same type of loading whatever the deflection, i.e. an orthogonal repartition of forces along the whole fiber length. This strong assumption is certainly no more valid in regime III when the fiber adopts a hook shape, that is when the angle $\theta(s=L)$ exceeds $\pi/2$.

\begin{figure}
\includegraphics[width=0.9\linewidth]{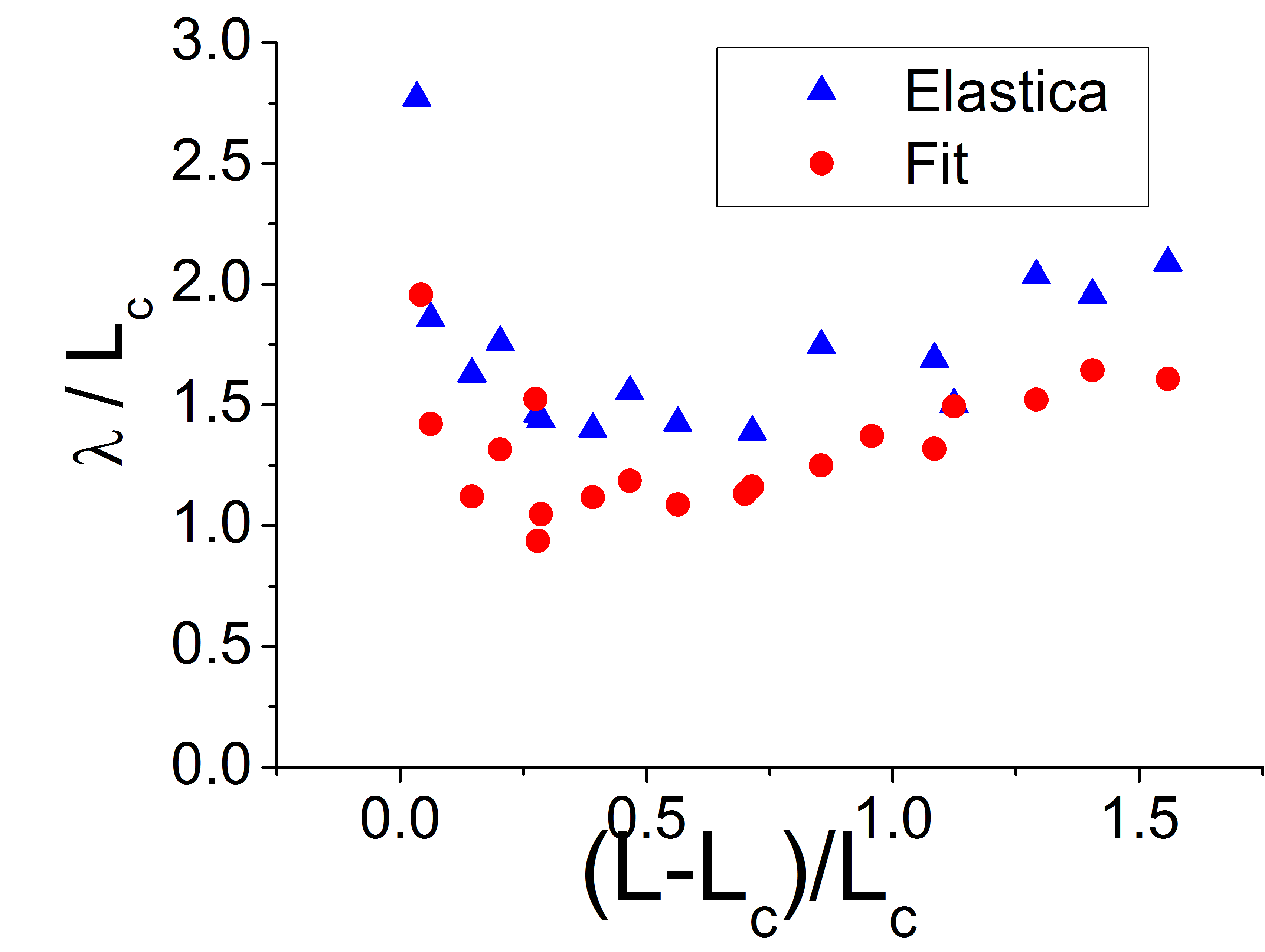}
 \caption{Characteristic length scale $\lambda$ for the rate of bending as a function of $\epsilon=\frac{L-L_c}{L_c}$. Red circles are $\lambda$ values extracted from the exponential fit of deflection $\delta$ vs penetration distance $\ell$. Blue triangles are derived from the $A$ values of the compaction-induced bending model coupled with the elastica simulations.}
  \label{LambdaLc}
\end{figure}

\subsection{Discussion of the model}
 
The values of $A$ that match the curves of deflection like in Fig.~\ref{Elastica-compaction} have been obtained for the different experiments performed with various fiber lengths $L$ or packing fractions $\phi$. For getting a physical meaning of $A$ in terms of length, we derive the corresponding $\lambda$ values from Eq.~\ref{lambda}. The $\lambda$ value characterizes the typical penetration distance necessary to bend the fiber in regime II once the bifurcation occurs. In Fig.~\ref{LambdaLc} these $\lambda$ values (blue triangles) are plotted as a function of the relative excess length above $L_c$, that is $\epsilon=\frac{L-L_c}{L_c}$. The values of $\lambda$ (red circles) obtained from the direct fit  $\delta=\delta_0 exp(\ell / \lambda)$ of Fig.~\ref{DeltaMoy} are also plotted on the same curve for comparison. For both derivations, the $\lambda$ values are of the order of $L_c$ but both curves also exhibit the same complex and non-monotonous evolution with $\epsilon$. For large $\epsilon$, that is for long fiber lengths, the $\lambda$ value seems to be governed by the excess length $L-L_c$. However for $\epsilon \rightarrow 0$, that is for fiber length approaching the critical length for bending, $\lambda$ increases by a factor of 2. For a fiber length just at the transition $L=L_c$ ($\epsilon = 0$) one would expect a divergence of $\lambda$, as $L_c$ provides the limit for the non-bending case. The increase of $\lambda$ observed for smaller $\epsilon$ might be interpreted in this framework. In any cases, the Fig.~\ref{LambdaLc} provides a intriguing relationship between $\lambda$, the length scale characterizing the rate of bending, and $L_c$ defining the critical fiber length for bending.

The elastica-derived model, though based on simple and minimal arguments, well captures the full range of the bending transition (regime II) (Fig.~\ref{Elastica-compaction}), even if it can not describe the deflection in regime III. More refined and complex loadings of the fiber (like non orthogonal forces or non homogeneous amplitudes or locations of point forces...)  might be investigated to improve the matching of the model with experimental deflections.

Moreover, even in the bending regime II, the observed non-monotonous evolution of $\lambda$ with $\epsilon$ probably indicates that there are two competing effects with increasing fiber lengths. Following the work of \cite{Holmes-EML16} on the buckling of elastic beams in granular media, it is tempting to consider that the fiber can be decoupled in two parts: one part of effective length $L_{eff}$ (probably related to $L_c$) that might supports the force exerted by the cluster, while the complementary part of length $L-L_{eff}$ would be simply led by the granular flow like a pinned rod. Direct measurements of forces and torques exerted on the fiber are needed to clearly identify the way the fiber is loaded. Complementary elastica simulations would provide a systematic way of quantifying the role of the fluctuating and discrete nature of force transmissions specific of granular material at the grain-fiber contacts. The approach to jamming introduces a further complex step in this new and intriguing fluid-structure interaction between and a flexible fiber and a dense granular flow.

\section{Conclusion\label{conclusion}}

This work presented a new fluid/structure interaction between a granular flow close to the jamming transition and a flexible fiber in a geometry of penetration. We identified a bending transition occuring for fiber longer than a characteristic length, that we called elasto-granular length $L_c$ as in the very recent work of \cite{Holmes-arxiv17}. In our case, the reconfiguration of the fiber shape was due to bending and not buckling but one can imagine to promote mechanisms of buckling by progressively increasing the aspect ratio between the fiber thickness and the grain diameter. 

The bending transition was associated with a symmetry breaking in the packing fraction, with the formation of a cluster upstream of the fiber and a cavity empty of grains downstream of the fiber. We proposed an expression for $L_c$ that combines both rigidities of the flexible fiber and granular material approaching the jamming transition. The elasto-granular length $L_c$ controlled the transition between the bending and jiggling regimes for the fibers embedded in the granular medium. But surprisingly $L_c$ was also observed to determine the kinetics of the bending once the transition started.

\section*{Acknowledgements}
We acknowledge the crucial technical help of T. Darnige as well as
enlightning discussions with P. Gondret and A. Seguin. We also acknowledge all the graduated students who have contributed to this work, Marguerite Leang, Yuka Takehara (from the group of K. Okumura), Francois Postic as well as Katherin Luginbuhl, Dawn Wendell and Nadia Cheng from the group of A. Hosoi through the MIT-France Seed Fund project on "Flexible Objects in Granular Media". 

\bibliography{granulence}

\end{document}